\documentclass[aip,graphicx]{revtex4-1}

\draft 

\usepackage{bm}
\usepackage{subcaption}
\usepackage[export]{adjustbox}
\usepackage{pbox}
\usepackage{makecell}
\usepackage{array}

\usepackage[dvipsnames]{xcolor}
\usepackage{colortbl}
\usepackage{graphicx}
\newcolumntype{P}[1]{>{\centering\arraybackslash}p{#1}}

\usepackage{float}

\begin{document}


\title{Dynamics of intracycle angular velocity control applied to cross-flow turbines} 



\author{Sara F. Hartke}
\email[]{sara.hartke@wisc.edu}
\affiliation{University of Wisconsin-Madison}
\author{Ari Athair}
\affiliation{University of Washington}
\author{Owen Williams}
\affiliation{University of Washington}
\author{Jennifer A. Franck}
\affiliation{University of Wisconsin-Madison}



\begin{abstract}
Understanding the intricate dynamics of cross-flow turbines (CFT) is critical to the improvement of performance and optimal control strategies. The current study numerically investigates intracycle control by modulating the angular velocity as a function of blade position for a 2-bladed NACA0018 turbine at a lab-scale chord-based Reynolds number of 45,000. Previous work has implemented intracycle control in attempts to improve turbine efficiency at the best performing tip-speed ratio (TSR). However, intracycle modulation of angular velocity simultaneously changes the time-averaged TSR, making it difficult to understand if the effects on performance are due to changes in mean TSR or imposed by the intracycle dynamics. Thus, this work explores a wider region of TSR across which intracycle control is applied, and assesses turbine performance with respect to time-averaged TSR. 
The effect of intracycle amplitude and phase shift of the velocity modulation is reported in terms of power generation and blade-level forces, and time-resolved flow fields reveal mechanisms behind changes in efficiency. 
For the 2-bladed turbine explored, the peak performance at constant angular velocity occurs at approximately TSR = 2. Intracycle control is found to be most beneficial at TSR $< 2$ where power is increased up to $71\%$ over its constant speed baseline and $12\%$ over the peak performance without control. This is accomplished through boundary layer reattachment through the acceleration portion of the stroke. In contrast, applying control when TSR $\ge 2$ is not beneficial due to degraded performance in the downstream portion of the stroke. 
\end{abstract}


\maketitle 

\section{Introduction}\vspace{-0.5cm}
This research computationally examines the intracycle flow physics of a cross-flow turbine (CFT) operated under non-uniform angular velocity. CFTs, also known as vertical axis wind turbines (VAWTs) when operated in air, are utilized as marine energy converters for fast-moving water currents. Distinct from the more common axial-flow turbines, their axis of rotation is perpendicular to the freestream velocity. Among other benefits, this feature allows for omnidirectional operation, reducing complexity by eliminating the need for active yaw control. 

\begin{figure}[htbp]
    \centering%
    \includegraphics[width=1\textwidth]{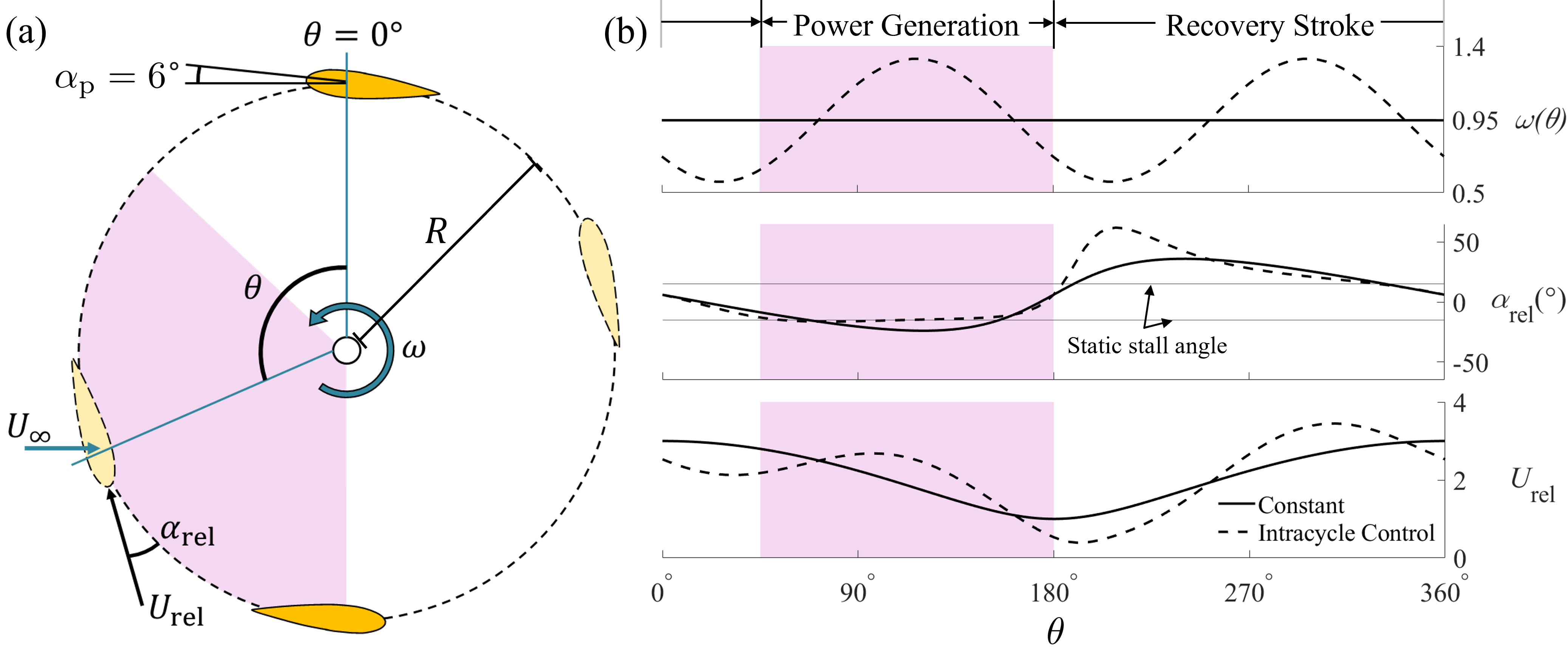}%
    \vspace{-0.25cm}%
    \caption{Diagram of cross-flow turbine blade (a) showing angular velocity, relative angle of attack, and relative blade velocity over one cycle for constant rotation and with intracycle control implemented (b).}%
    \label{fig:turbineDiagram}%
    \vspace{-0.5cm}%
\end{figure}

Depicted in Figure \ref{fig:turbineDiagram}a is the geometry of the CFT cross-section utilized in this study with two straight blades. Turbine blade position is defined by phase, $ \theta $, throughout a cycle, with $\theta=0^\circ$ as the position of the blade when it is traveling directly upstream. For a two-bladed turbine, each blade has the same cycle, offset by $ 180 ^{\circ} $. The rotation rate is characterized by the tip-speed ratio (TSR), which is the ratio of the tangential velocity of the blade to the freestream velocity, 
\begin{equation}
    \lambda(\theta) = \frac{\omega(\theta) R}{U_{\infty}},
    \label{eq:TSR}
\end{equation}

\noindent where $\omega$ is the turbine angular velocity, $R$ is the turbine radius, and $U_\infty$ is the freestream velocity. Despite their geometric simplicity, CFT blades encounter dynamic changes in angle of attack ($\alpha_{\text{rel}}$) and relative velocity ($U_{\text{rel}}$) during rotation (Fig. \ref{fig:turbineDiagram}b), initiating complex flow physics that intrinsically link operational kinematics and turbine geometry with turbine performance. 
This paper explores controlling these blade-level physics by sinusoidally varying the angular velocity over a turbine cycle, resulting in improved power extraction.

Two distinct approaches can be used to optimize the performance of CFTs. The first focuses on alterations to the blades and/or turbine geometry, such as the number of blades, turbine solidity, blade chord to turbine radius ratio, or fixed blade pitch angle, all of which affect the optimal rotation rate of the turbine and overall performance \cite{rezaeiha2018,duarte2022,kumar2022,araya2017, hunt2024}. These geometrical changes are also strongly dependent on Reynolds number \cite{hunt2024}, which changes the flow regime in which turbines operate, and are further complicated by variation of relative flow speed and angle of attack of the blade. 

Previous research has investigated CFT dynamics at the blade level to help inform the mechanisms leading to power improvements \cite{snortland2019, dave2023,lefouest2022}. 
Over one cycle, a CFT blade experiences a range of angles of attack (Fig. \ref{fig:turbineDiagram}b), which commonly exceed the static stall angle \cite{allet1997,mandal1994}. 
Soon after the static stall angle is exceeded, a vortex forms on the suction side of the blade until it separates and sheds in a process known as dynamic stall, which initiates an abrupt loss of lift \cite{ferreira2009,lefouest2022}. At the blade level, decreasing separation and delaying dynamic stall improves power generation as it holds the low pressure vortex core close to the blade surface for longer \cite{amet2009}.
Equally important, strong and abrupt dynamic stall cycles induce oscillations in forces on the blade and turbine structure, which can reduce lifespan or cause turbines to be over-designed at high cost \cite{tsang2008}. 
Thus, the interplay between dynamic stall and performance is essential, and should be considered in the design and operation of CFTs. For example, at lower TSR the dynamic stall vortex often sheds dramatically and prematurely, before the natural change in angle of attack would precipitate flow separation. Rather than simply operate at higher rotational speeds, on-blade control can be utilized to actively or passively change blade-level dynamics \cite{zhao2022}. For instance, the addition of vortex generators can reduce boundary layer separation and increase lift, and adding dimples on the blade surface can enhance torque output, but these methods add complexity and cost \cite{choudhry2016}. 

A second category for CFT performance improvement is the optimization and control of turbine kinematics, which is the focus of the current investigation. The primary strategies explored in previous studies are the intracycle control of blade pitch angle \cite{kirke2011,paraschivoiu2009,schonborn2007,gauthier2017,zhang2023,lefouest2024} or angular velocity \cite{strom2017,dave_aiaa_2021,athair2023}, both of which directly alter the relative angle of attack of the blade and thus change the dynamic stall behavior and downstream recovery. 
Blade pitch control yields improvements to self-start capability and reduction of unsteady forces \cite{kirke2011}, a weak point for CFTs, and higher power generation at lower TSR was found using cyclic pitching of a straight-bladed Darrieus turbine \cite{schonborn2007}. 
Through exploration of sinusoidal and arbitrary pitch control, optimal settings to modulate dynamic pitch angle have demonstrated up to approximately $30\%$ improvement in efficiency \cite{paraschivoiu2009}. 
Though performance improvements are observed under optimal settings, the additional monetary and energy cost of added motors to change the blade pitch may outweigh the potential benefits \cite{strom2017,schonborn2007}. 

Intracycle control of angular velocity has the promise to deliver similar benefits without requiring additional hardware. 
Studies varying angular velocity throughout the turbine cycle have found improvement in performance, both in power generation and unsteady force loading \cite{strom2017,dave_aiaa_2021,dave2023,athair2023}. These investigations control angular speed as a function of blade position using variations of a sinusoidal function, 
\begin{equation}
    \lambda(\theta) = \lambda_\theta + A' \sin(N\theta + \phi'),
    \label{eq:control_orig}
\end{equation}
\noindent where the angular velocity oscillates about the phase-averaged TSR, $\lambda_\theta$, defined as 
\begin{equation}
    \lambda_\theta = \frac{1}{2\pi}\int\limits_0^{2\pi} \lambda(\theta) \, \mathrm{d}{\theta},
    \label{eq:phase-avg-tsr}%
\end{equation}

\noindent at prescribed amplitude $A'$ and phase shift $\phi'$, where $N$ is the number of turbine blades. The free parameters $A'$ and $\phi'$ determine the magnitude of velocity oscillation and where in the blade cycle the sinusoidal acceleration and deceleration occur. 

For a 2-bladed turbine, \citet{strom2017} performed an optimization procedure on $\lambda_\theta$, $A'$, and $\phi'$ to assess power improvement of intracycle control with respect to operation at constant TSR. Results reported power improvement up to $59\%$ at $\lambda_{\theta}=1.9$, the optimal constant TSR for the turbine \cite{strom2017}, finding little difference between sinusoidal versus a more generalized parameterization of angular velocity control. However, at $Re_{\text{c}}=31,000$, the benefits of intracycle control were compounded by low Reynolds number effects and boundary layer transition and so the overall improvement in power output was  overestimated to some extent. Computationally, the flow physics have been explored on a similar turbine geometry using both unsteady Reynolds-averaged Navier-Stokes (RANS) \cite{dave_aiaa_2021} and large eddy simulation (LES) \cite{dave2023}. The improvements in power through intracycle velocity control were compared with constant angular velocity at equivalent phase-averaged TSR and ascribed to a delay in boundary layer separation, and thus a delay in the onset of dynamic stall. These positive increases in power, up to $40\%$, occur at specific values of $\phi'$ where the peak intracycle velocity aligns with the peak torque generation \cite{dave2023}. A follow-up experimental investigation at $Re_{\text{c}} = 44,000$ shows a reduction in peak turbine loading of up to $12\%$, along with power improvement up to $14\%$ applying control near the optimal constant speed TSR \cite{athair2023}. In each of these previous investigations, phase shift greatly affected the impact of control on performance. Exploring a range of phase shifts is necessary because under intracycle control, the acceleration and deceleration elicit different blade-level flow dynamics depending on where in the cycle they occur \cite{athair2023}. 

The previous exploration of intracycle control has demonstrated the effectiveness of manipulating the dynamic stall vortex to improve power output. However, there are natural gaps of knowledge that this paper and subsequent research seek to address. All previous studies have focused on implementation of control at the optimal TSR for constant angular velocity. At a non-optimal TSR, the flow separation and dynamic stall process is detrimental, perhaps leading to more opportunity for improvement with control. Secondly, besides the work by \citet{athair2023}, there has been little exploration of the amplitude of intracycle control, nor its impacts on the flow physics. 
Finally, prior research has assessed the benefits of control by measuring against the equivalent phase-averaged TSR at constant angular velocity. However, efficiency of the turbine is measured in terms of the time-averaged power coefficient, thus an argument can be made that intracycle angular velocity control should be compared against an equivalent time-averaged TSR for the turbine at constant angular velocity. 

Thus, this paper computationally investigates the kinematic and dynamic consequences of intracycle velocity control over a wide range of TSRs, amplitudes, and phase shift angles. Using a 2D RANS model, simulations are validated with complementary lab-scale experimental data. The kinematics of the angular motion are carefully defined with time-averaged TSR values, offering a more consistent comparison against baseline constant velocity conditions. Finally, the performance differences in power and forces are computed, and the flow physics for representative kinematics are discussed in detail. 
\vfill
\section{Methods}
\label{sec:methods}\vspace{-0.5cm}

\subsection{Turbine and Flow Parameters}\vspace{-0.5cm}
\label{subsec:turbParameters}
The turbine parameters in the simulations include two straight NACA0018 blades at a preset pitch of 6 degrees (radially outwards) about the quarter chord (Fig. \ref{fig:turbineDiagram}a) and a turbine radius $ R = 2.118c $, where $c$ is the blade chord length. The chord-based Reynolds number is $ Re_{\text{c}} = {cU_\infty}/{\nu} = 45,000$ and the blockage ratio is 10.6\%, determined by the projected area of the turbine over total cross-sectional area of the domain, both of which are determined by the capabilities of the experimental testing facilities and experimental conditions in \citet{athair2023}. Results for this blockage ratio are considered close to that of an unconfined turbine \cite{dave_aiaa_2021} and are uncorrected for blockage because the investigated TSRs are low enough that dynamic stall occurs; thus the power extracted is relatively insensitive to blockage \cite{kinsey2017}.

Power is nondimensionalized as 
\begin{equation}
    C_{P}(\theta) =  \frac{q(\theta) \omega(\theta)}{\frac{1}{2} \rho U_{\infty}^3 A },
    \label{eq:powerCoeff}
\end{equation}
\noindent where $q$ is torque, $\rho$ is fluid density, and $A$ is turbine projected area. Most commonly $\omega$ is a constant for cross-flow turbines, but here it is an arbitrary function of $\theta$ to allow for intracycle velocity control. 

Additionally, streamwise and cross-stream forces on each turbine blade are nondimensionalized as
\begin{equation}
    C_{X}(\theta) = \frac{F_{X}(\theta)}{\frac{1}{2} \rho U_{\infty}^2 A }
\end{equation}
and 
\begin{equation}
    C_{Y}(\theta) = \frac{F_{Y}(\theta)}{\frac{1}{2} \rho U_{\infty}^2 A }, 
\end{equation}
\noindent respectively. 

To compute the power coefficient, the nondimensional power is integrated with respect to time over one period, $ T $, 
\begin{equation}
    \overline{C_P}(t) = \int\limits_0^T C_P(t) \, \mathrm{dt}.  
    \label{eq:avgPowerCoeff_time}
\end{equation}

\subsection{Intracycle Velocity Control Equations}\vspace{-0.5cm}
Equation \ref{eq:control_orig} defines the sinusoidal velocity about a phase-averaged TSR (Eq. \ref{eq:phase-avg-tsr}). An alternative but equivalent approach is 
\begin{equation}
    \lambda(\theta) = \lambda_\theta + A_\lambda\overline{\lambda} \cos{N(\theta-\phi)}, 
    \label{eq:intracycleNew}
\end{equation}
which introduces a time-averaged TSR,
\begin{equation}
    \overline{\lambda} = \frac{1}{T} \int\limits_0^T \lambda \, \mathrm{dt},
    \label{eq:time-avg-tsr}%
\end{equation}
\noindent where $A_\lambda$ is the amplitude of the velocity oscillation as a percentage of the time-averaged TSR and $\phi$ is the phase shift. Note that $ \phi $ is related to $ \phi' $ in Equation \ref{eq:control_orig} by $\phi=45^\circ - \phi'/N$ and is a more natural choice since $\phi$ directly identifies where in the cycle angular velocity is highest, consistent with the expression used in \citet{athair2023}. For example, at $\phi=117^\circ$, the highest angular velocity occurs at $\theta=117^\circ$ and is periodic with the number of turbine blades. Since the angular velocity is not  constant, the position of the blade as a function of time, $\theta(t)$, is the solution to the following equation formed from Equation \ref{eq:TSR},

\begin{equation}
    \frac{\lambda(\theta) U_{\infty}}{R} = \frac{\mathrm{d \theta}}{ \mathrm{dt}},
    \label{eq:dthetadt}%
\end{equation}

\noindent and is computed numerically.

\begin{figure}[htbp]
    \centering%
    \includegraphics[width=1\textwidth]{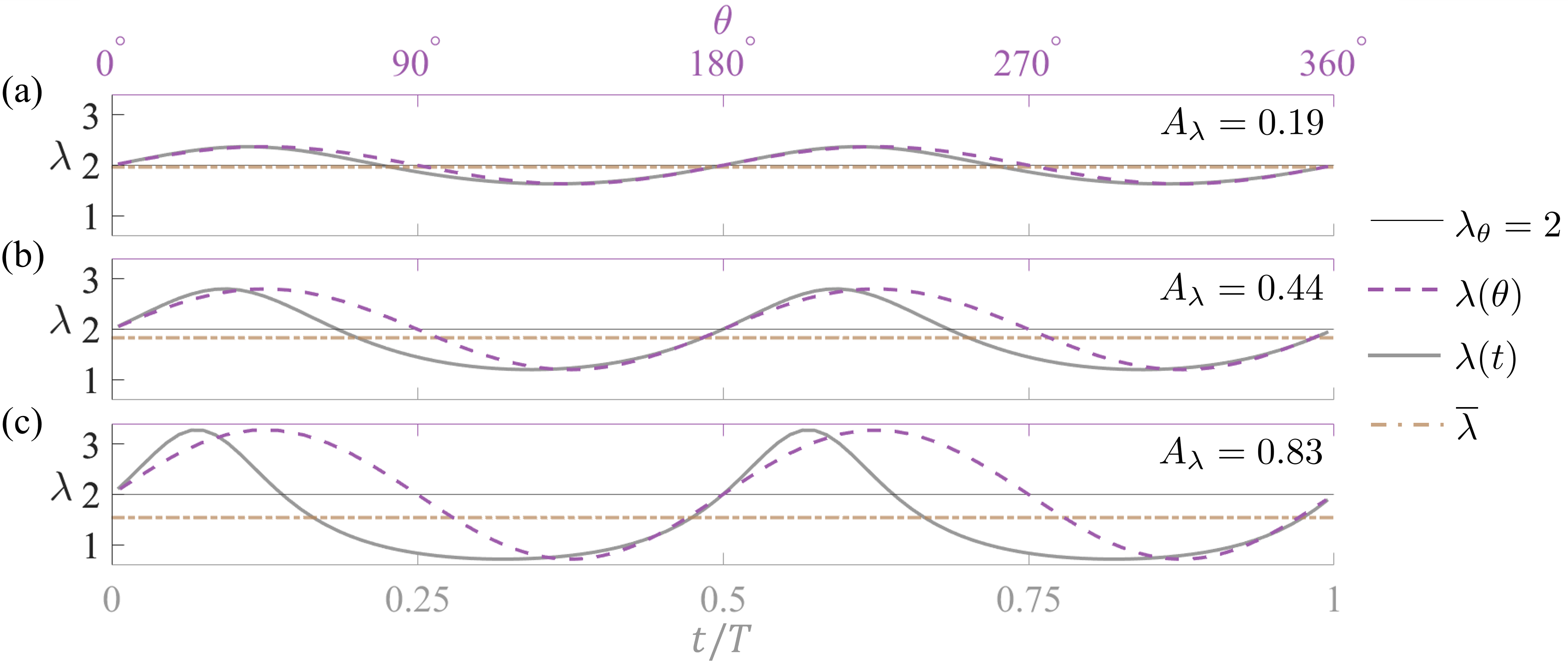}%
    \vspace{-0.25cm}%
    \caption{Tip-speed ratio versus phase and time (normalized by the period) for different amplitudes of intracycle velocity variation at $\phi=45^\circ$.}%
    \label{fig:tsrCurves}%
    \vspace{-0.25cm}%
\end{figure}

When implementing intracycle velocity control, comparisons are made with respect to a baseline (uncontrolled) state at constant angular velocity with the same TSR as the intracycle mean. 
It is unclear, however, whether the intracycle mean  should be phase-averaged or time-averaged. 
Figure \ref{fig:tsrCurves} demonstrates how the angular velocity varies as a function of phase and time with constant {\it phase-averaged} TSR. 

Some notable differences emerge when TSR is plotted versus phase (top x-axis), as shown by the purple dashed line, and versus time (bottom x-axis), as shown by the solid grey line. The intracycle control variables are defined to be sinusoidal with respect to phase (Equation \ref{eq:intracycleNew}), and TSR increases in amplitude without losing symmetry, as expected, with a mean of $ \lambda_\theta = 2 $. However, the angular velocity with respect to time is not symmetric about a TSR of 2. As the amplitude of oscillation increases, a distinct asymmetric curve develops, with considerably greater time being spent at lower angular velocity (Fig. \ref{fig:tsrCurves}c). Thus, averaging over time reveals a changing mean TSR ($\overline{\lambda}$), shown by the tan dot-dashed lines in Figure \ref{fig:tsrCurves}, that is lower than the phase-averaged value $\lambda_\theta$. This change is amplified with increasing $A_{\lambda}$. Thus, when previous studies explored changes to turbine performance due to intracycle at a constant phase-averaged TSR, the time-averaged TSR was lowering as amplitude increased. 

These results indicate that changes in power output for angular velocity control can be due to 1) a shift in mean TSR with respect to time and 2) intracycle acceleration and deceleration of the blade. Comparisons of equal time-averaged TSR would isolate the second cause. Thus, this manuscript proposes intracycle simulations at equivalent time-averaged TSRs to assess performance in Section \ref{subsec:effTSR} while seeking opportunities to augment turbine performance.  

\vspace{-0.75cm}
\subsection{Numerical Methods}\vspace{-0.5cm}
\subsubsection{Computational Solver and Domain}\vspace{-0.5cm}
The simulations utilize open-source \textit{OpenFOAM} libraries to solve the incompressible RANS equations,
\begin{equation}
    \nabla \cdot \boldsymbol{ u } = 0
    \label{eq:NS1_mass}
\end{equation}
and
\begin{equation}
    \frac{\partial \boldsymbol{u}}{\partial t} + (\boldsymbol{ u } \cdot \nabla ) \boldsymbol{ u } = - \nabla p + \nu \nabla ^2 \boldsymbol{ u } - \nabla \cdot \boldsymbol{\tau} ,
    \label{eq:NS2_momentum}
\end{equation}%
\noindent where $ \boldsymbol{u } $ is the Reynolds-averaged velocity vector, $ p $ is the Reynolds averaged pressure, and $\nu$ is the kinematic viscosity. A $ k\text{-}\omega $ SST turbulence model approximates the undetermined terms of the Reynolds stress tensor, $ \boldsymbol{\tau} $. This turbulence model has been demonstrated to work well for RANS simulations of turbines \cite{buchner2015,balduzzi2016}. The solver discretizes the equations spatially with a second-order finite volume method and temporally with a second-order implicit method before using the pressure-implicit, split-operator (PISO) algorithm. For more details on the RANS implementation, see \citet{dave_jrse_2021}. 

The full computational domain has a size of $450c$ in the streamwise and $40c$ in the cross-stream direction. The turbine is placed at a location centered transversely and $200c$ downstream of the inlet boundary within an AMI sliding interface of diameter $8c$ (Fig. \ref{fig:meshDetails}). The boundary conditions on the inlet are constant velocity and a zero-gradient pressure, whereas on the outlet they are zero-gradient velocity and a fixed pressure. On the top and bottom surfaces, a slip boundary is prescribed, whereas the airfoil surface has no-slip boundary conditions. 

To impose the non-uniform motion, angular velocity is calculated for each phase and constant time steps are prescribed with corresponding positions of turbine blades. These data are put into an array which dynamically rotates the turbine according to each time-phase data point. For each TSR, a sweep of 3 amplitudes, $A_\lambda$, and 5 phase shifts, $ \phi $, between $ 0\text{ - }180^{\circ} $ are simulated to investigate effects of intracycle velocity variations throughout the blade cycle. The reported power is phase-averaged over the last five cycles after simulations reach a fully developed state.

\vspace{-0.75cm}
\subsubsection{Mesh Resolution}
\begin{figure}[htbp]
    \centering%
    \begin{subfigure}{0.31\textwidth}
        \subcaption{}%
        \includegraphics[width=1\textwidth]{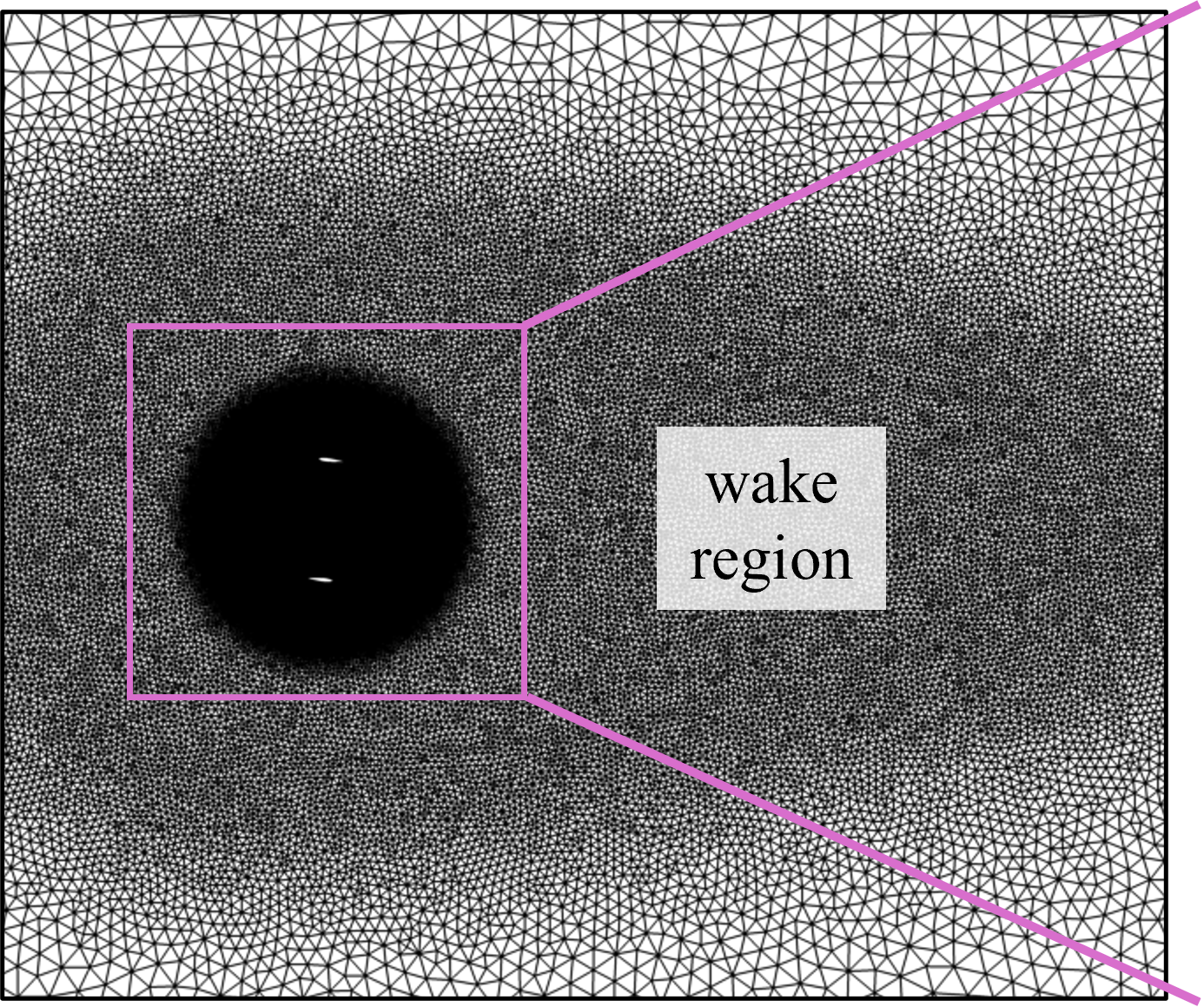}%
        \label{fig:expCp_a}%
    \end{subfigure}%
    \begin{subfigure}{0.31\textwidth}
        \subcaption{}%
        \includegraphics[width=1\textwidth]{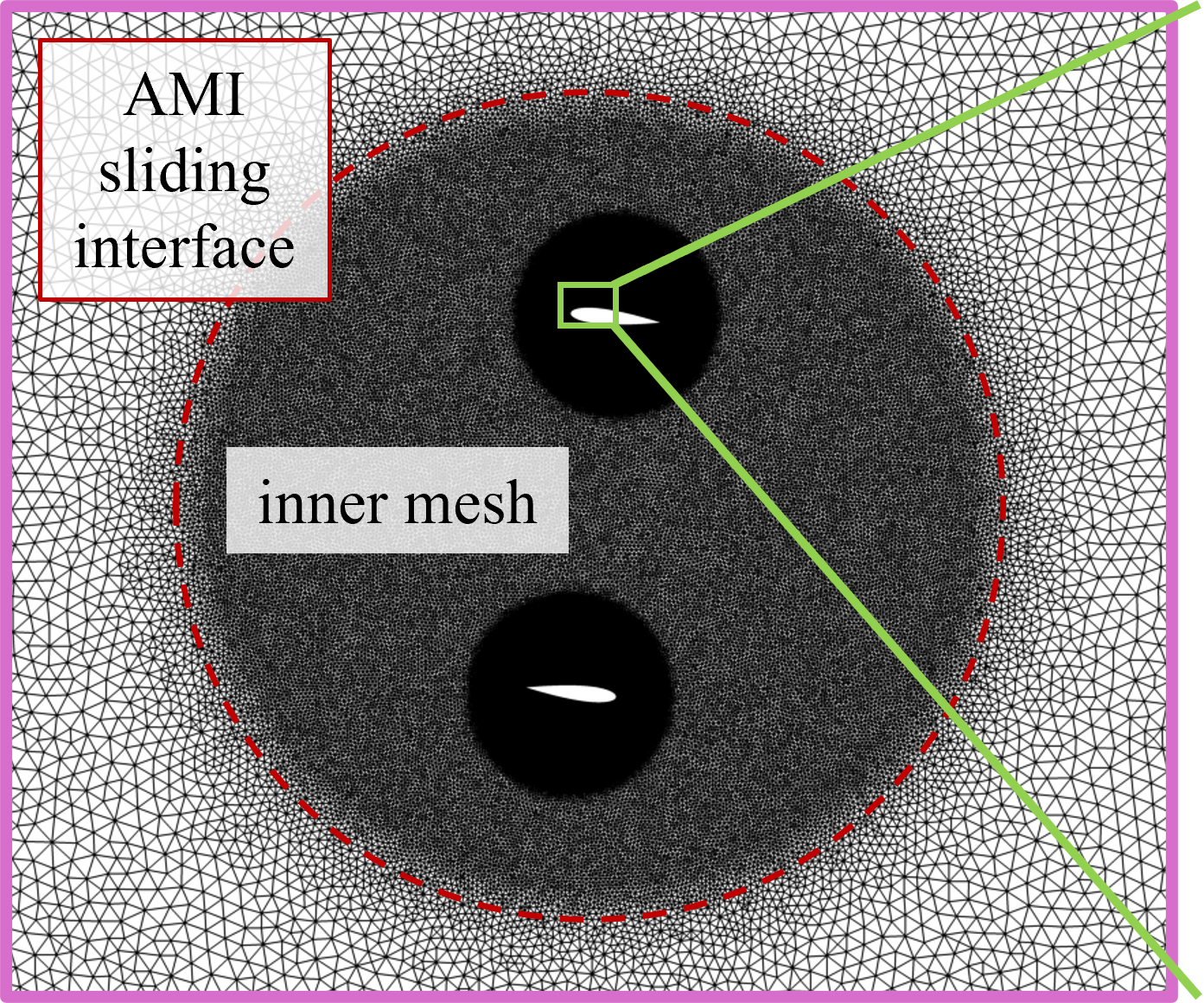}%
        \label{fig:expCp_a}%
    \end{subfigure}%
    \begin{subfigure}{0.305\textwidth}
        \subcaption{}%
        \includegraphics[width=1\textwidth]{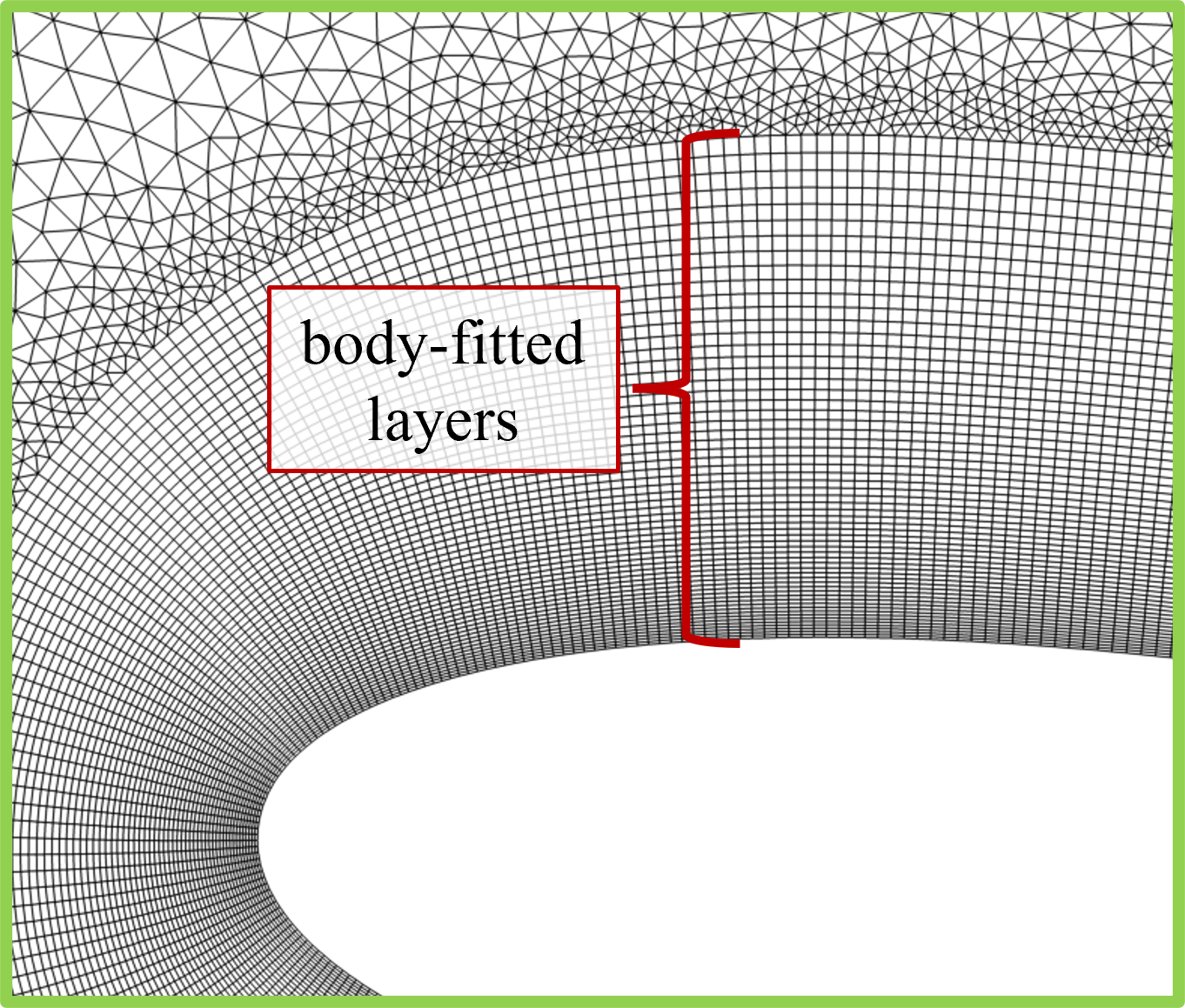}%
        \label{fig:expCp_a}%
    \end{subfigure}%
    \vspace{-0.25cm}%
    \caption{Mesh layers at different resolution identifying areas changed in mesh study.}%
    \label{fig:meshDetails}%
    \vspace{-0.25cm}%
\end{figure}

\vspace{-0.5cm}
The mesh resolution is similar to previously validated RANS turbine simulations \cite{dave_aiaa_2021,dave_jrse_2021}. However, given the rapid changes in velocity with implementation of intracycle velocity control, mesh resolutions are repeated here under sinusoidal velocity control, and summarized in Table \ref{tab:meshConvergence}. The data demonstrates that blade-level resolution has the most significant effect on resulting power. The mesh utilized in simulations is Mesh 1a, pictured in Figure \ref{fig:meshDetails}. 

Increasing resolution in the wake region (Fig. \ref{fig:meshDetails}a) as done in Meshes 1a-c results in the same power coefficient. Compared with Mesh 1a, Mesh 2a increases the resolution in the inner mesh, increasing the number of cells by 80\%, but changes the power coefficient by less than 0.1\%. With the same number of body-fitted layers, Mesh 2a and 2b examine the effect of the height of the first body-fitted layer ($\Delta y/c$). The resulting $11.5\%$ difference between power coefficients indicates that further consideration is needed to determine which resolution is most accurate using the $k\text{-}\omega \textnormal{ SST}$ turbulence closure model, since further resolution does not guarantee convergence for the RANS equations. 
Compared with Mesh 2b, Mesh 3 reduces the number of body-fitted layers from 60 to 50, which results in a power difference of $0.5\%$, indicating that $50$ body-fitted layers are sufficient. Additionally, it points to the change in power coefficient noted in comparing meshes 2a and 2b being primarily due to changes in the height of the first layer rather than changes in total height of the entire body-fitted layer region. 
To address the $\Delta y /c $ resolution, comparisons are made for constant TSR and intracycle control between mesh 1a, mesh 3, and experimental results, in which it is found that both meshes produce results well-matched with the experiments; however, mesh 1a more closely predicts the constant TSR power curves shown by experimental results and thus is selected for use in this research.

\begin{table}[htbp] 
    \caption{Mesh configuration details checking mesh sensitivity for RANS intracycle simulations at $ \lambda_\theta = 2; A_\lambda = 0.44; \phi = 117^{\circ} $}.%
    \label{tab:meshConvergence}%
    \centering%
    \vspace{-0.5cm}%
    \begin{tabular}{l|P{25mm}|P{30mm}|P{20mm}|P{15mm}|P{15mm}}
        & $ \Delta y/c $ of first mesh layer & \# of body-fitted layers & \# of inner mesh cells & Total \# of cells & $ \overline{C_{\text{P}}} $ \\ \hline 
        mesh 1a     & 0.001 & 60  & 178,077 & 272,892 & 0.3489\\ 
        mesh 1b     & 0.001 & 60  & 178,077 & 282,412 & 0.3489\\ 
        mesh 1c     & 0.001 & 60  & 178,077 & 292,184 & 0.3489\\ 
        mesh 2a     & 0.001 & 60  & 318,837 & 400,646 & 0.3487\\
        mesh 2b     & 0.0005 & 60  & 316,017 & 420,352 & 0.3888\\ 
        mesh 3     & 0.0005 & 50  & 185,209 & 280,024 & 0.3890\\ 
    \end{tabular}%
    \vspace{-0.25cm}%
\end{table}

\vspace{-0.5cm}
\subsection{Experimental Methods}\vspace{-0.5cm}
Parallel experiments are conducted at the Alice C. Tyler recirculating water flume at the University of Washington. Turbine geometry parameters match those specified in Section \ref{subsec:turbParameters}, with a dimensional blade chord length of $c = 4.06$ cm. 
The turbine has a blade span of 0.234 m, which corresponds to an aspect ratio of 1.36. Support struts with NACA0008 profile are used to mount the blades at their ends to minimize drag. Test conditions are held constant and match the simulation's non-dimensional parameters. A pool heater and chiller are used to maintain constant temperature ($\pm 0.1 ^\circ C$) and thus a constant chord-based Reynolds number, $Re_{\text{c}} = 44,000$. The dynamic depth (57.2 cm) and inflow velocity (0.885 m/s) are also fixed to maintain a constant blockage ratio of 9.4\% and submergence-based Froude number $Fr_{\text{S}}=U_\infty/\sqrt{g S} = 0.67$, where g is the gravitational constant and $S$ is the distance from the surface to the top of the turbine. 

The turbine is controlled by a 5 N$\cdot$m servo motor (Yaskawa SGMCS-05B3C41) that has an internal encoder with 2$^{18}$ counts per revolution resolution that records the motor position and velocity throughout the cycle. The intracycle velocity profile is prescribed in \textit{MATLAB Simulink} and actuated by the motor. Two six-axis ATI load cells (Mini45) are mounted at the top and bottom of the turbine assembly to 
measure the forces and torques on the turbine while sweeping through a range of intracycle control parameters. Additional details about the experimental setup can be found in \citet{hunt2024}. A test time of 45 seconds is used for each intracycle kinematic tested, allowing for the elimination of any cycle-to-cycle variation. Both control of the motor and data acquisition is conducted at a 1 kHz rate through a National instruments data acquisition system. Simultaneously, an acoustic Doppler velocimeter, Nortek Vectrino is used to collect velocity measurements undisturbed by the turbine structure for use in performance non-dimensionalization. It measures mean flow velocity at 16 Hz, $10 R$ upstream of the turbine at the mid span of the blades. 

Two sets of supplementary experiments are conducted to isolate the blade-level performance, necessary for comparison to simulations. First, tests are conducted under matching experimental conditions with blades removed. These allow support losses to be estimated and subtracted, isolating blade-only hydrodynamic measurements following the method of \citet{strom2018SuportStruct}. Second, the turbine is swept through an arbitrary intracycle profile in air to estimate the moment of inertia. This is used to remove inertial loading from the phase-averaged data, following the method developed by \citet{strom2017} and combined with the support subtraction by \citet{athair2023}.
All performance data is post-processed using a Butterworth low-pass filter with 
a cutoff frequency of 100 Hz and 10 Hz pass frequency to filter out high-frequency electromagnetic interference. This filter has been found to have no impact on time-averaged performance data by \citet{snortland2025downstream}. The filtered data is trimmed to an integer number of rotations before either time- or phase-averaging, allowing for comparison with simulations. 

In addition to performance measurements, flow field data are acquired using particle image velocimetry (PIV) at the turbine mid span for three intracycle kinematics at $\lambda_\theta=2$. These PIV measurements utilize a similar experimental setup. The same turbine is used; however, it is configured to be a cantilevered system with a large acrylic plate on the bottom so that a high-speed camera (Phantom V641) can capture the in-rotor PIV from a position beneath the flume. This camera has a resolution of  2560$\times$1600. A 30 mJ/pulse Continuum Terra PIV Nd:YLF laser illuminates the flow in a 2 mm thick horizontal sheet. Hollow glass neutrally-buoyant particles with a 10 $\mu$m diameter are used to seed the flow. Two fields-of-view of 15.7$\times$25.1 cm$^2$ are collected, one in the upstream region and one in the downstream with a 3 cm overlap. The time between frames is set to 350-370 $\mu$s such that particle displacements are small enough to resolve the vortex cores. PIV timing is synchronized with the turbine blade position such that 85 image pairs are collected at each phase, allowing for phase-averaging of the results for higher data quality. Data are collected for approximately every 12 degrees of the blade rotation for a total of 30 phases. 

Raw images are post-processed using \textit{Lavision DaVis 10.2}. A moving-average background subtraction is applied to the raw images in batches of 49 images to remove any constant reflections and noise present. Manual geometric masks are drawn for each phase to remove the blade and shadowed regions. Multi-pass PIV calculation is conducted with 5 passes of interrogation windows starting with 128$\times$128 pixels down to 32$\times$32 pixels, with 75\% overlap. Vector validation post-processing is applied using a universal outlier detection method on a filter domain of 5$\times$5 vectors with a threshold of three standard deviations. No interpolation or smoothing of data is performed. After all processing, the resulting vector fields are scaled to dimensional values from a calibration factor of 10.16 pixels/mm and exported into \textit{MATLAB} for further analysis and plotting.

\section{Results}
\label{sec:results}\vspace{-0.5cm}

Results are divided into three sections. First, a sequence of simulations with and without control closely matched to experiments are analyzed at constant {\it phase-averaged TSR}, $ \lambda_\theta = 2 $. Second, simulations with angular velocity control at constant {\it time-averaged TSR}, $\overline{\lambda} = 2 $ are compared. Finally, the influence of departures in the time-averaged TSR away from the optimal values found for constant rotation rates is investigated, allowing the identification of regions where intracycle control could be beneficial to CFT operation. 

\vspace{-0.75cm}
\subsection{Comparison with experiments: constant phase-averaged TSR}
\label{subsec:validation}\vspace{-0.5cm}
A sequence of 17 simulations are presented with varying $ A_\lambda $ and $ \phi$ at $ \lambda_\theta = 2 $. These are selected to match experimental conditions in \citet{athair2023} for comparison and validation. Table \ref{tab:expComp} summarizes the resulting time-averaged power coefficients (Eq. \ref{eq:avgPowerCoeff_time}) for each set of computational and experimental kinematics. Overall, only moderate improvement in performance is achieved with control at constant $\lambda_\theta=2$. 

\begin{table}[htbp] 
    \caption{Comparison of power coefficient $\overline{C_{\text{P}}}$ at $\lambda_\theta = 2 $ produced by simulations and experiments with constant angular velocity and with intracycle angular velocity control. Cell colors indicate percent difference in power under intracycle control with respect to the baseline (constant TSR) at each $\overline{\lambda}$.}%
    \vspace{-0.25cm}%
    \label{tab:expComp}%
    \centering%
    \begin{tabular}{l||P{18mm}|P{11mm}P{11mm}P{11mm}|P{18mm}|P{11mm}P{11mm}P{11mm}}
        & \multicolumn{4}{c|}{Computational} & \multicolumn{4}{c}{Experimental} \\ \hline
        $  $ & Constant & \multicolumn{3}{c|}{Intracycle} & Constant & \multicolumn{3}{c}{Intracycle} \\ \hline 
        $ \overline{\lambda} $  & 2 & 1.97  & 1.83 & 1.54 & 2 & 1.97  & 1.83 & 1.54 \\
        $ A_{\lambda} $        & 0 & 0.19  & 0.44 & 0.83 & 0 & 0.19  & 0.44 & 0.83 \\ \hline
        $ \phi=9^{\circ} $    & 0.336 & \cellcolor{orange!14}0.312  & \cellcolor{orange!100}0.164 & \cellcolor{orange!100}0.054 & 0.360 & \cellcolor{orange!33}0.300 & \cellcolor{orange!77}0.221 & \cellcolor{orange!100}0.110 \\ 
        $ \phi=27^{\circ} $    & - & -  & \cellcolor{orange!81}0.201 & - & - &  \cellcolor{orange!21}0.322 & \cellcolor{orange!55}0.261 & \cellcolor{orange!100}0.171 \\ 
        $ \phi=45^{\circ} $     & - & \cellcolor{orange!20}0.302  & \cellcolor{orange!37}0.274 & \cellcolor{orange!64}0.228 & - & \cellcolor{orange!18}0.328 & \cellcolor{orange!39}0.289 & \cellcolor{orange!59}0.254 \\
        $ \phi=81^{\circ} $   & - & \cellcolor{violet!01}0.338  & \cellcolor{violet!03}0.340 & \cellcolor{orange!11}0.318 & - &  \cellcolor{orange!02}0.356 & \cellcolor{orange!03}0.355 & \cellcolor{orange!12}0.338 \\
        $ \phi=117^{\circ} $   & - & \cellcolor{violet!11}0.354  & \cellcolor{violet!08}0.349 & \cellcolor{violet!27}0.381 & - & \cellcolor{violet!11}0.380 & \cellcolor{violet!20}0.396 & \cellcolor{violet!23}0.401 \\ 
        $ \phi=153^{\circ} $    & - & \cellcolor{violet!07}0.347  & \cellcolor{orange!87}0.189 & \cellcolor{orange!100}0.018 & - & \cellcolor{violet!06}0.370 & \cellcolor{orange!23}0.319 & \cellcolor{orange!100}0.067 \\
        $ \phi=171^{\circ} $    & - & -  & \cellcolor{orange!100}0.158 & - & - &  \cellcolor{orange!16}0.331 & \cellcolor{orange!87}0.204 & \cellcolor{orange!100}0.069 \\ \hline%
        \multicolumn{9}{c}{\includegraphics[width=0.5\textwidth]{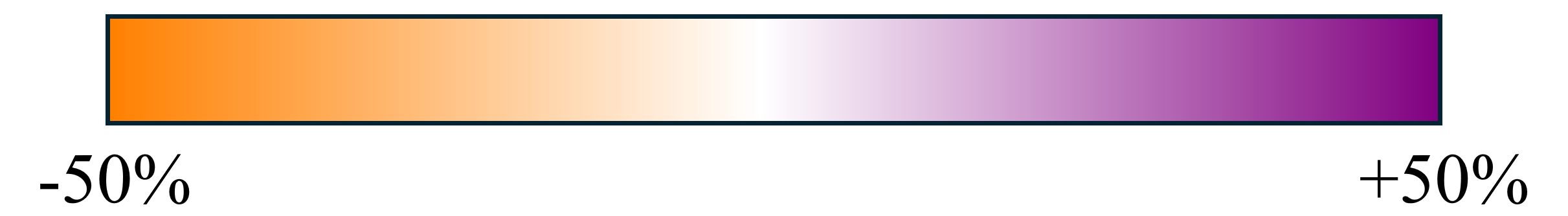}}
    \end{tabular}%
    \vspace{-0.5cm}
\end{table}
\vspace{-0.5cm}
Table \ref{tab:expComp} demonstrates good correspondence in power coefficients between numerical and experimental results, with a $6.7\%$ difference under constant speed (baseline), and similar trends across all control parameters, reporting percent differences from baseline within 3\% in over half of the cases and less than 15\% for all others, the notable exception being at $A_\lambda=0.44; \phi=153^\circ$ where simulations predict 32\% more power loss than experimental data indicates. Note that more phase shifts are computed for $A_\lambda=0.44$ to compare with the finer resolution of $\phi$ in experimental data. 

Both experiments and computations show a strong dependence on phase shift angle, with $ \phi = 9^{\circ} $ resulting in decreased power output at all amplitudes and $ \phi = 117^{\circ} $ giving the highest power coefficient at all amplitudes. 
At $\phi=117^\circ$, improvements in power are observed across all $A_\lambda$, from $5.6\text{ - }11.4\%$ experimentally and $5.3\text{ - }13.4\%$ numerically at the low and high amplitudes, respectively. 
This aligns with the dynamics of a turbine, where for $ \phi = 9^{\circ} $, any intracycle control will have the blades moving the slowest during peak torque generation and continuing to be slow after blade stall. At $A_\lambda=0.44$, both datasets agree that $\phi=171^\circ$ performs the worst, with $-43\%$ in experiments and $-53\%$ in simulations. 
Simulations also capture the more moderate power differences at distinct phase shifts, showing $\pm1\%$ difference in power at $\phi=81^\circ$ and $19\text{ - }20\%$ difference at $\phi=45^\circ$ between simulations and experiments.

\begin{figure}[htbp]
    \vspace{-0.25cm}%
    \centering%
    \begin{subfigure}{0.3936\textwidth}
        \subcaption{Power versus phase}%
        \includegraphics[width=1\textwidth]{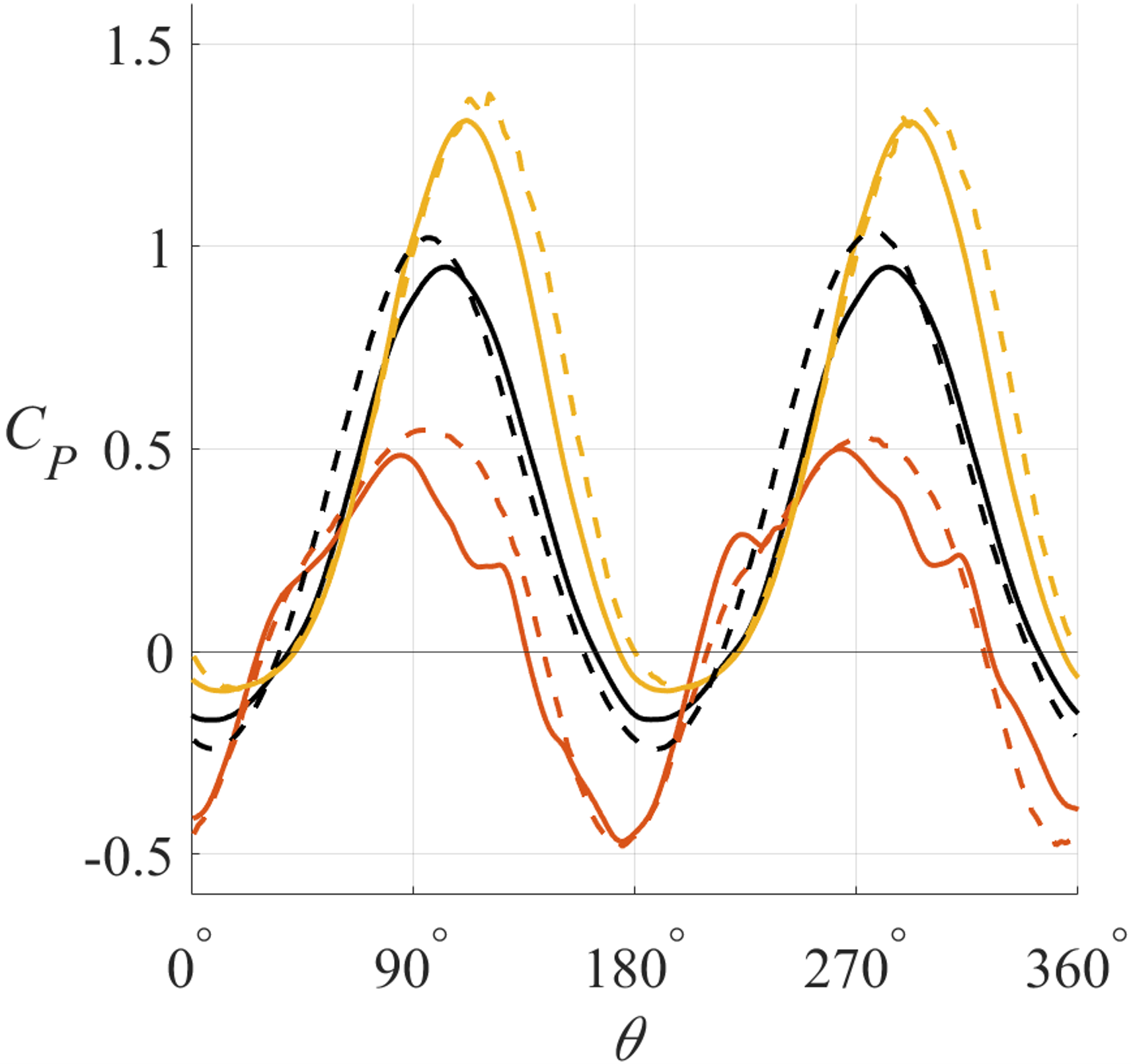}%
        \label{fig:expCp_a}%
    \end{subfigure}%
    \begin{subfigure}{0.5288\textwidth}
        \subcaption{Power versus time}%
        \includegraphics[width=1\textwidth]{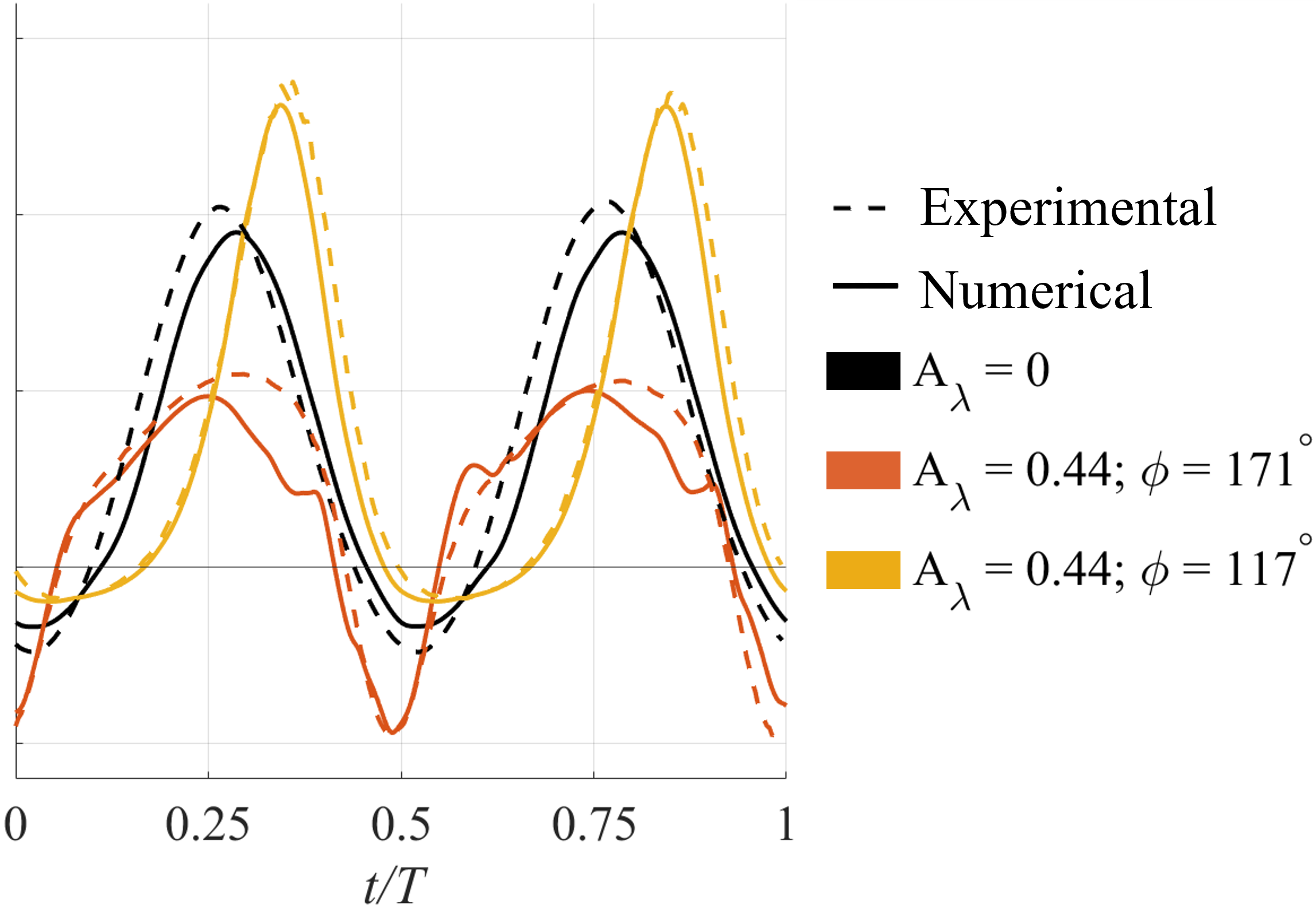}%
        \label{fig:expCp_b}%
    \end{subfigure}%
    \vspace{-0.25cm}%
    \caption{Power comparison between simulations and experiments at $\lambda_\theta = 2$ with respect to phase (a) and time (b), at a constant rotation rate and for intracycle control at $ A_\lambda = 0.44 $ and sub-optimal and optimal phase shifts.}%
    \label{fig:expComparison_cp}%
    \vspace{-0.25cm}%
\end{figure}
\begin{figure}[htbp]
    \centering%
    \begin{subfigure}{0.5\textwidth}
        \subcaption{$\theta \approx 36^{\circ}$}%
        \includegraphics[width=1\textwidth]{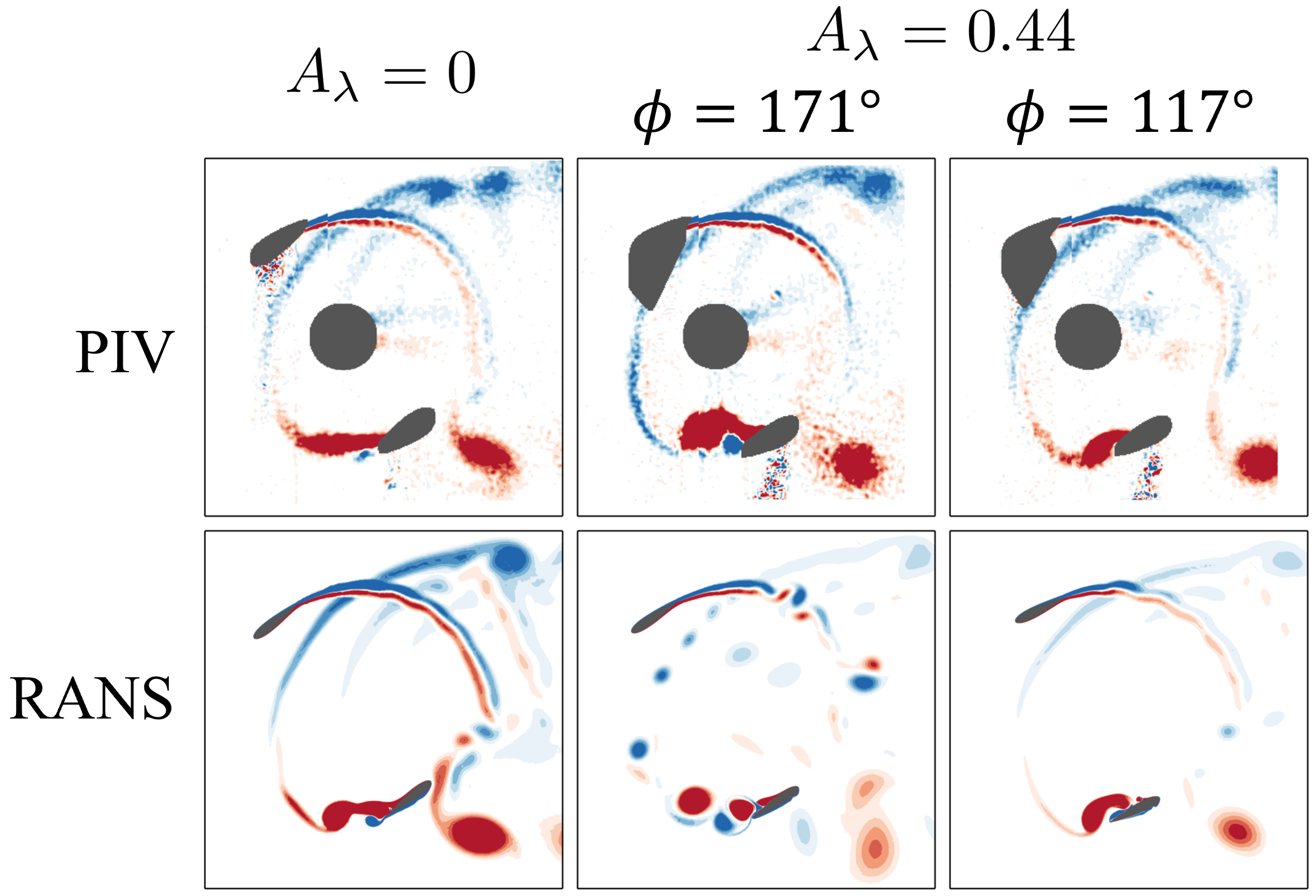}%
        \label{fig:3_1_a_vort}%
    \end{subfigure}
    \begin{subfigure}{0.5\textwidth}
        \subcaption{$\theta \approx 72^{\circ}$}%
        \includegraphics[width=1\textwidth]{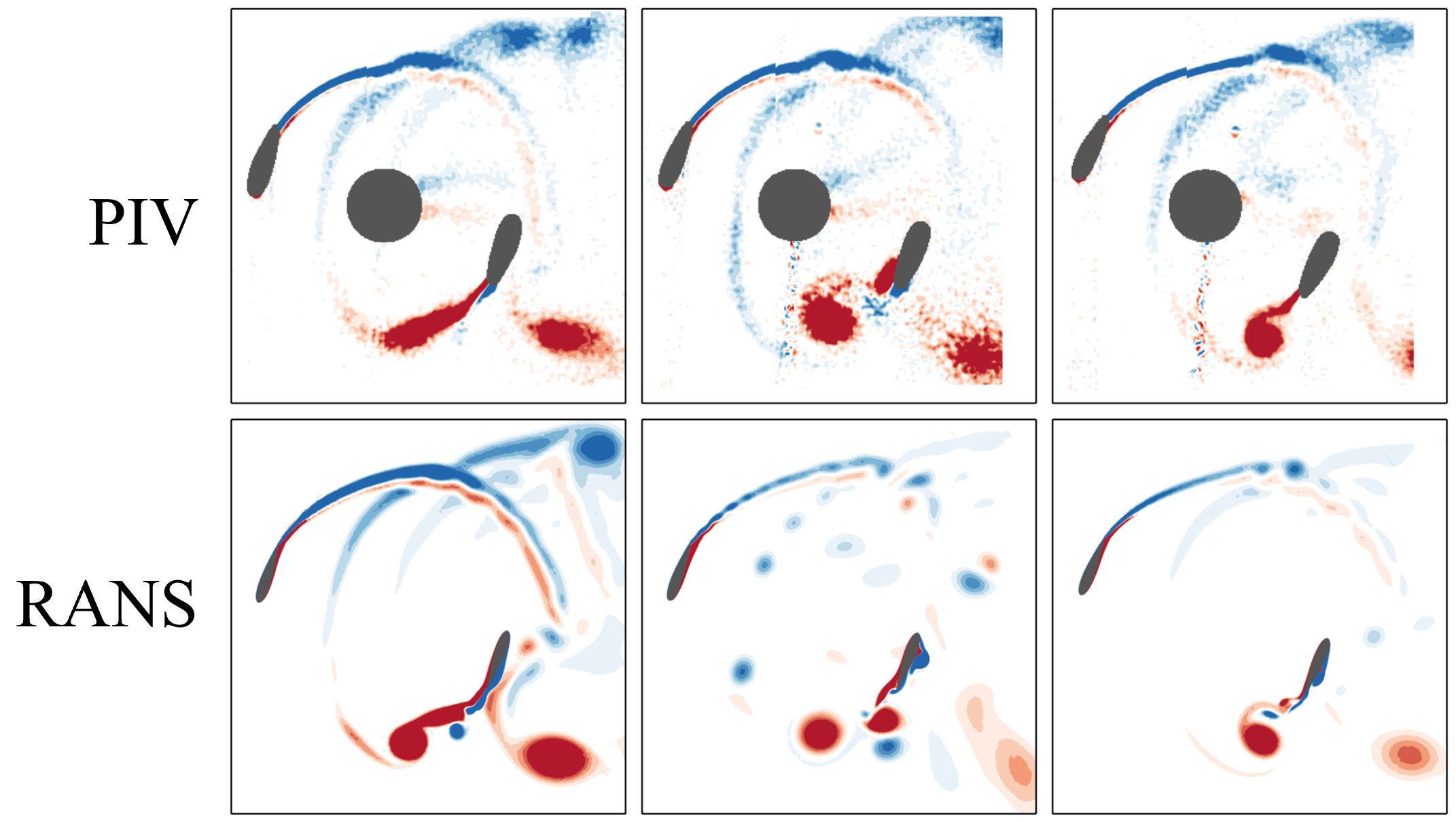}%
        \label{fig:3_1_b_vort}%
    \end{subfigure}
        \begin{subfigure}{0.5\textwidth}%
        \subcaption{$\theta \approx 120^{\circ}$}%
        \includegraphics[width=1\textwidth]{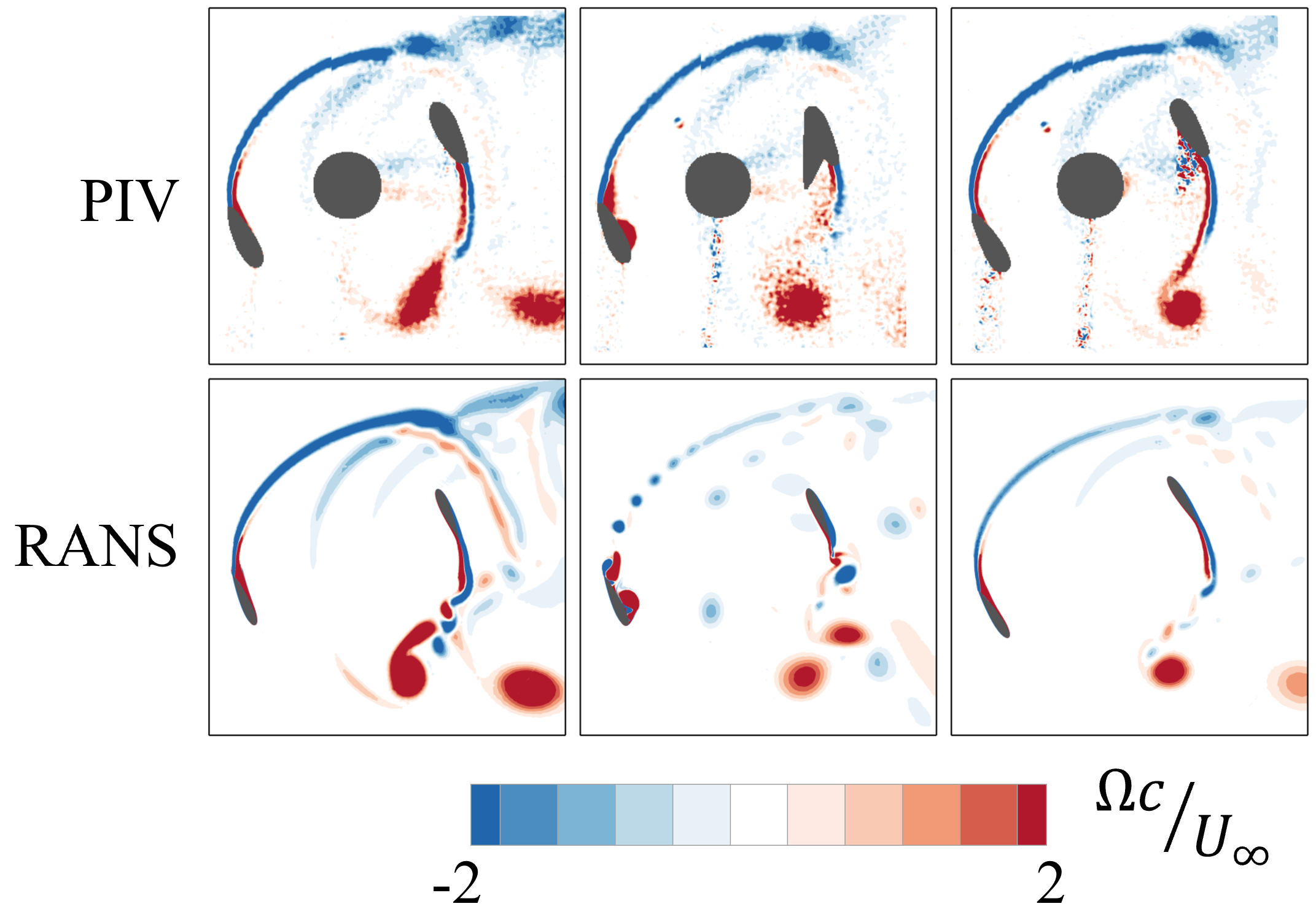}%
        \label{fig:3_1_c_vort}%
    \end{subfigure}
    \vspace{-0.5cm}%
    \caption{Spanwise vorticity comparison between experimental PIV and computational RANS at $\lambda_\theta = 2$. }%
    \label{fig:expComparison_vort}%
    \vspace{-0.5cm}%
\end{figure}

The time-dependent power curves are compared between computational and experimental datasets in Figures \ref{fig:expComparison_cp} and \ref{fig:expComparison_vort}. Power can be explored in either a phase ($\theta$) or time ($t/T$) basis, with differences between the curves growing with intracycle amplitude as shown in Figure \ref{fig:expComparison_cp}, which compares the baseline, constant rotation rate case, as well as optimal and sub-optimal intracycle phases. When plotted with respect to phase, the width of the curve in the first half of the cycle is similar between baseline (constant TSR) and optimal control ($ A_\lambda = 0.44, \phi = 117^\circ $). However, when plotted with respect to time, the power curve for optimal control narrows over the power stroke, unveiling little overall integrated improvement in power coefficient when compared with baseline. If comparisons are made by averaging power over phase, the greater area under the curve would misleadingly suggest significant improvement using this combination of control parameters. To evaluate the benefits of intracycle control, integrating power with phase and time are not equivalent. Time integration is the meaningful metric of power reported here (Eq. \ref{eq:avgPowerCoeff_time}).

Blade-level flow dynamics can be explanatory of the performance changes seen above and are compared for the same cases in Figure \ref{fig:expComparison_vort} (contours of instantaneous spanwise vorticity, $\Omega$) with PIV and RANS referring to experimental and numerical results, respectively. Overall, the features identified in both computational and experimental flow fields compare very favorably, with convection of detached vortices appearing similar in both the constant and intracycle controlled rotation at $ A_\lambda = 0.44; \phi = 117^\circ $. Under a sub-optimal phase shift of $ \phi=171^\circ $, the on-blade dynamics match between PIV and RANS, but the trailing vortices are more distinct in the RANS simulations. This has been previously reported in RANS CFT simulations and could be an artifact of the RANS equations \cite{dave_aiaa_2021, dave_jrse_2021}. 

As in prior works \cite{rezaeiha2018,athair2023}, the optimal TSR for this turbine at constant angular velocity is $\lambda=2$, where the dynamic stall and separation of the trailing edge vortex from the blades are delayed until the end of the power stroke. This characteristic appears most clearly at $ \theta = 120^\circ $ (Fig. \ref{fig:3_1_c_vort}) for a constant rotation rate and intracycle control with $A_\lambda=0.44; \phi=117^{\circ}$, where flow is nicely attached to the leading blade as it approaches the bottom of the turbine cycle. For the same amplitude, but different phase shift $\phi$, strong vortices are developing on the blade during the power stroke, reflected in both the PIV and RANS visualizations. Controlling their strength and timing of separation leads to the large differences in performance observed for the different control kinematics.

\vspace{-0.85cm}
\subsection{Simulations at constant time-averaged TSR}
\label{subsec:effTSR}\vspace{-0.5cm}
As shown previously in Figure \ref{fig:tsrCurves}, the time-averaged TSR differs from the phase-averaged TSR for any given intracycle control kinematics. The difference between phase- and time-averaged TSR is a function of the amplitude of sinusoidal velocity oscillation and increases with $ A_\lambda $. Here, results for simulations at constant time-averaged TSR ($\overline{\lambda}$) are computed. The motivation for this additional set of computations is to isolate the effect of the acceleration/deceleration of the blade by eliminating changes in time-averaged TSR seen in Section \ref{subsec:validation} (Table \ref{tab:expComp}). Keeping the time-averaged TSR constant while varying amplitude allows for a fairer comparison of flow physics and  power production across amplitudes of intracycle control. 

Table \ref{tab:netpower_effConstant} summarizes the power coefficient over one turbine cycle for each set of parameters simulated at $ \overline{\lambda}=2 $. The data demonstrates different trends than those observed in Table \ref{tab:expComp}, namely that any power generation improvements to be made through use of intracycle control are limited to the lowest amplitudes simulated. Control using higher amplitudes yields no significant improvements to the power coefficient at baseline. Similar trends with the phase shift are noted, though the beneficial range of phase shifts moves higher; thus, acceleration occurring later in the power stroke of the blade performs better. 

\begin{figure}[htbp]
    \centering%
    \includegraphics[width=0.75\textwidth]{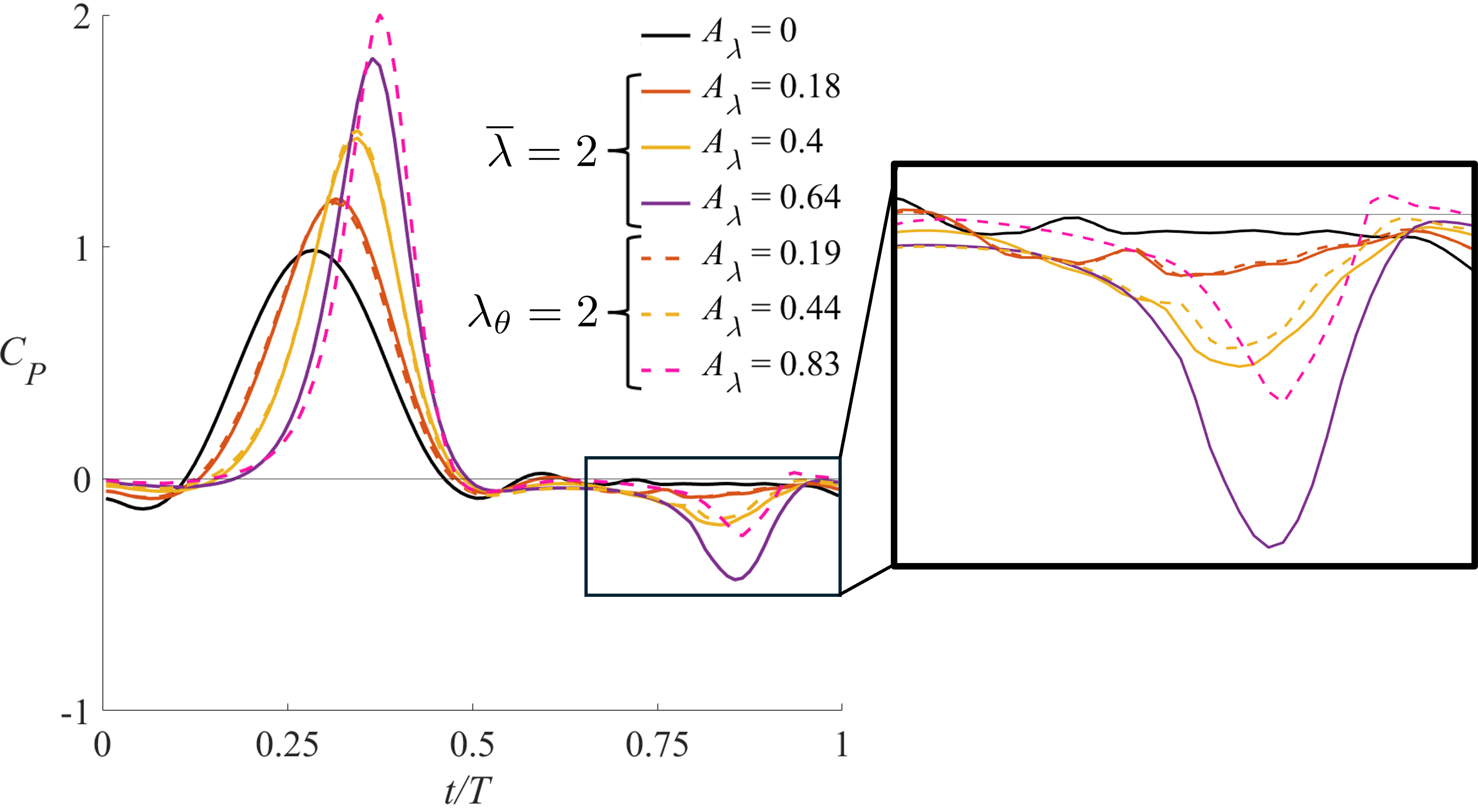}%
    \vspace{-0.25cm}%
    \caption{Power curves at $ \phi = 117^{\circ} $ for $\lambda_\theta = 2$ and $\overline{\lambda} = 2$ with implementation of low-, medium-, and high-amplitude intracycle control.}%
    \label{fig:constEff_powerCurve}%
    \vspace{-0.25cm}%
\end{figure}

\begin{table}[htbp] 
    \caption{Power coefficients $\overline{C_{\text{P}}}$ of simulations at $\overline{\lambda} = 2 $. Cell colors indicate percent difference in power compared with constant rotation rate.}%
    \label{tab:netpower_effConstant}%
    \vspace{-0.25cm}%
    \centering%
    \begin{tabular}{l||P{20mm}|P{14mm}P{14mm}P{14mm}}
        $  $ & Constant & \multicolumn{3}{c}{Intracycle} \\ \hline 
        $ \lambda_\theta $  & 2 & 2.04  & 2.16 & 2.38 \\
        $ A_{\lambda} $       & 0 & 0.18  & 0.4 & 0.64 \\ \hline
        $ \phi=9^{\circ} $   & 0.336 & \cellcolor{orange!06}0.326  & \cellcolor{orange!72}0.215 & \cellcolor{orange!100}0.119 \\ 
        $ \phi=45^{\circ} $    & - & \cellcolor{orange!08}0.321  & \cellcolor{orange!38}0.271 & \cellcolor{orange!70}0.220 \\
        $ \phi=81^{\circ} $  & -     & \cellcolor{violet!02}0.338  & \cellcolor{orange!20}0.304 & \cellcolor{orange!42}0.267 \\
        $ \phi=117^{\circ} $  & -     & \cellcolor{violet!14}0.358  & \cellcolor{violet!02}0.338 & \cellcolor{orange!20}0.302 \\ 
        $ \phi=153^{\circ} $   & -     & \cellcolor{violet!14}0.359  & \cellcolor{violet!04}0.342 & \cellcolor{orange!38}0.272 \\ \hline%
        \multicolumn{5}{c}{\includegraphics[width=0.5\textwidth]{figures/Cp_tableLegend_50.png}}
    \end{tabular}%
    \vspace{-0.5cm}%
\end{table}
\vspace{-0.25cm}
Power curves are compared between constant time- and phase-averaged TSR kinematics in Figure \ref{fig:constEff_powerCurve} for $\phi=117^\circ$. Similar trends in power emerge, with the peak power increasing as the amplitude of velocity control increases from low-to-high values by $67.9\%$ at $\lambda_\theta=2$ and by $50.3\%$ at $\overline{\lambda}=2$. Power loss in the wake appears to be of similar magnitude for both $\lambda_\theta=2$ and $\overline{\lambda}=2$ at low-to-mid $A_\lambda$. 
Interestingly, the downstream power losses for the highest amplitude $ \overline{\lambda}=2 $ are 76.8\% greater than those for $ \lambda_\theta =2$, despite having a lower intracycle amplitude. 

\begin{figure}[htbp]
    \centering%
    \includegraphics[width=0.5\textwidth]{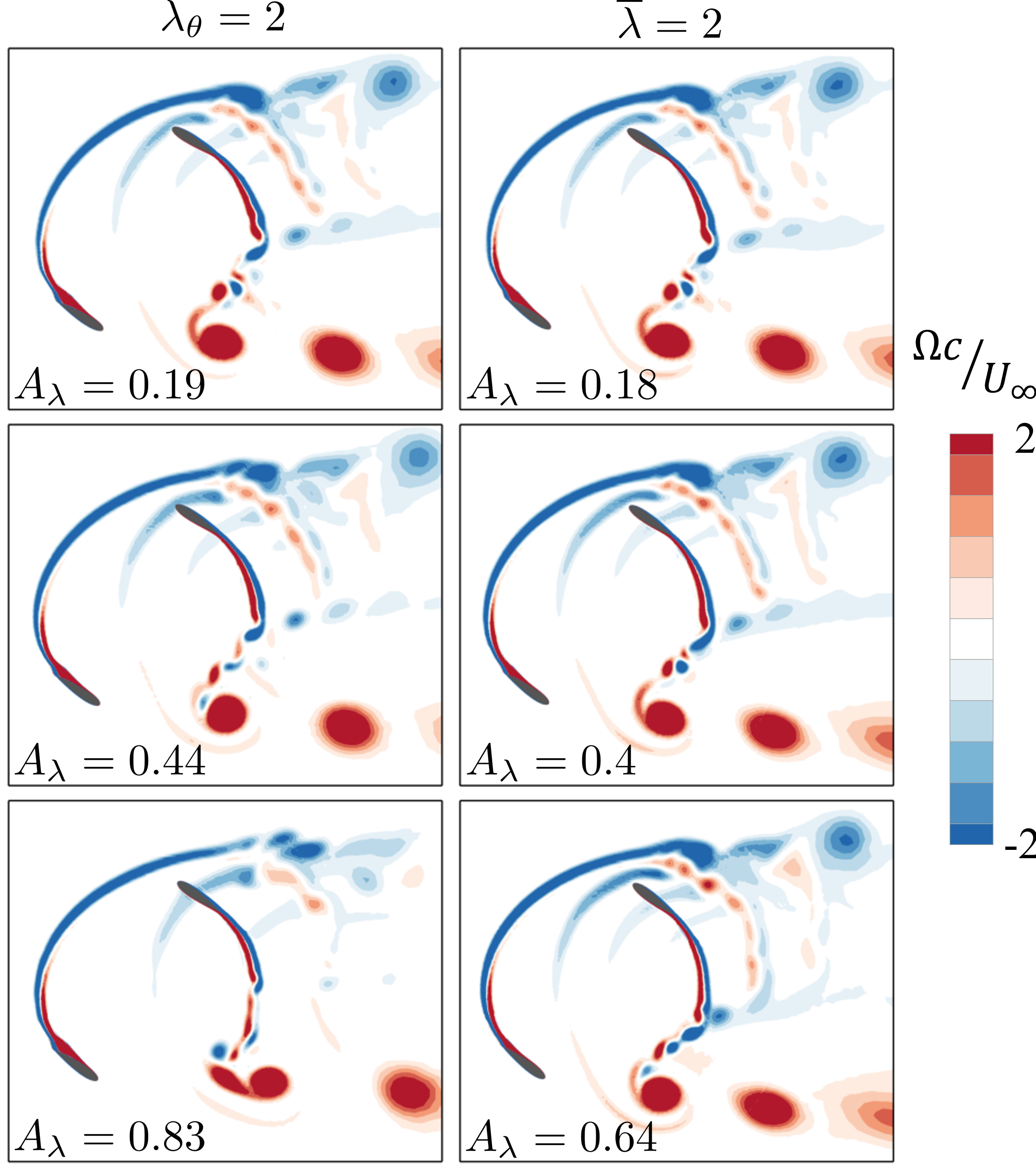}%
    \vspace{-0.25cm}%
    \caption{Spanwise vorticity comparison at $ \phi = 117^{\circ} $ between $\lambda_\theta = 2$ and $\overline{\lambda} = 2$ with implementation of low-, medium-, and high-amplitude intracycle control.}%
    \label{fig:compNomEff_vorticity}%
    \vspace{-0.25cm}%
\end{figure}

Instantaneous spanwise vorticity contours are shown in Figure \ref{fig:compNomEff_vorticity}, highlighting distinct differences at $\lambda_\theta=2$ and $\overline{\lambda}=2$ with intracycle control applied. Perhaps most apparent is that the dynamic stall vortices are shed with different timing, resulting in distinct wake patterns between the left and right columns. At low- and mid-amplitude control, the spacing and strength of vorticity is very similar, suggesting that the wake interactions on the downstream stroke are also comparable across time-averaged and phase-averaged TSRs, consistent with downstream power comparisons in Fig. \ref{fig:constEff_powerCurve}. However, at high-amplitude, the trailing blade of the time-averaged TSR passes through stronger vortices that appear closer together than the phase-averaged TSR, helping to explain observed differences in downstream power losses, as the blade interaction with these closer-spaced vortices is likely detrimental. 

\begin{figure}[htbp]
    \centering%
    \begin{subfigure}{0.48\textwidth}
        \subcaption{}%
        \vspace{-0.25cm}%
        \includegraphics[width=1\textwidth]{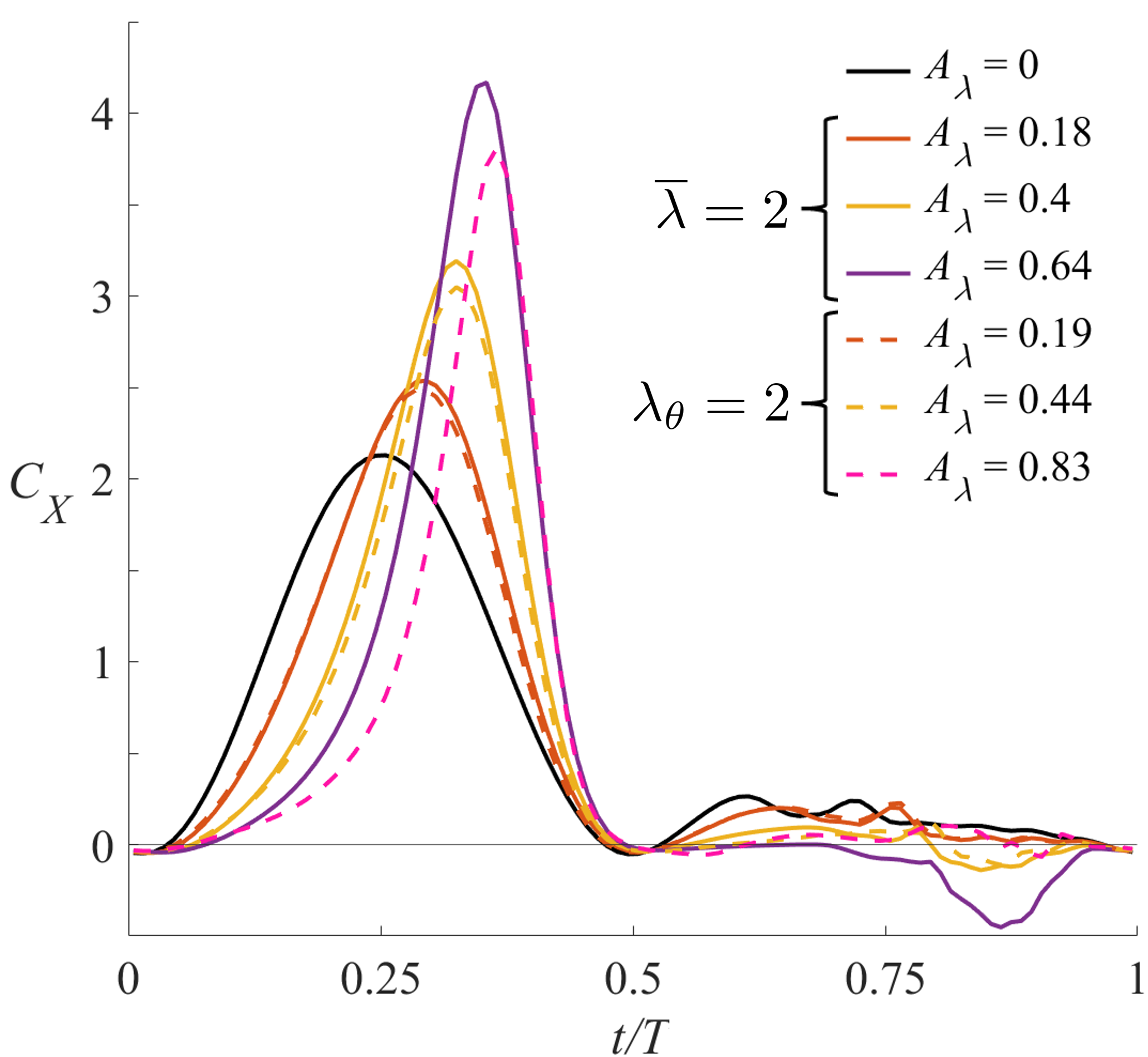}%
        \label{fig:3_2_a_forces}%
    \end{subfigure}
    \begin{subfigure}{0.48\textwidth}
        \subcaption{}%
        \vspace{-0.25cm}%
        \includegraphics[width=1\textwidth]{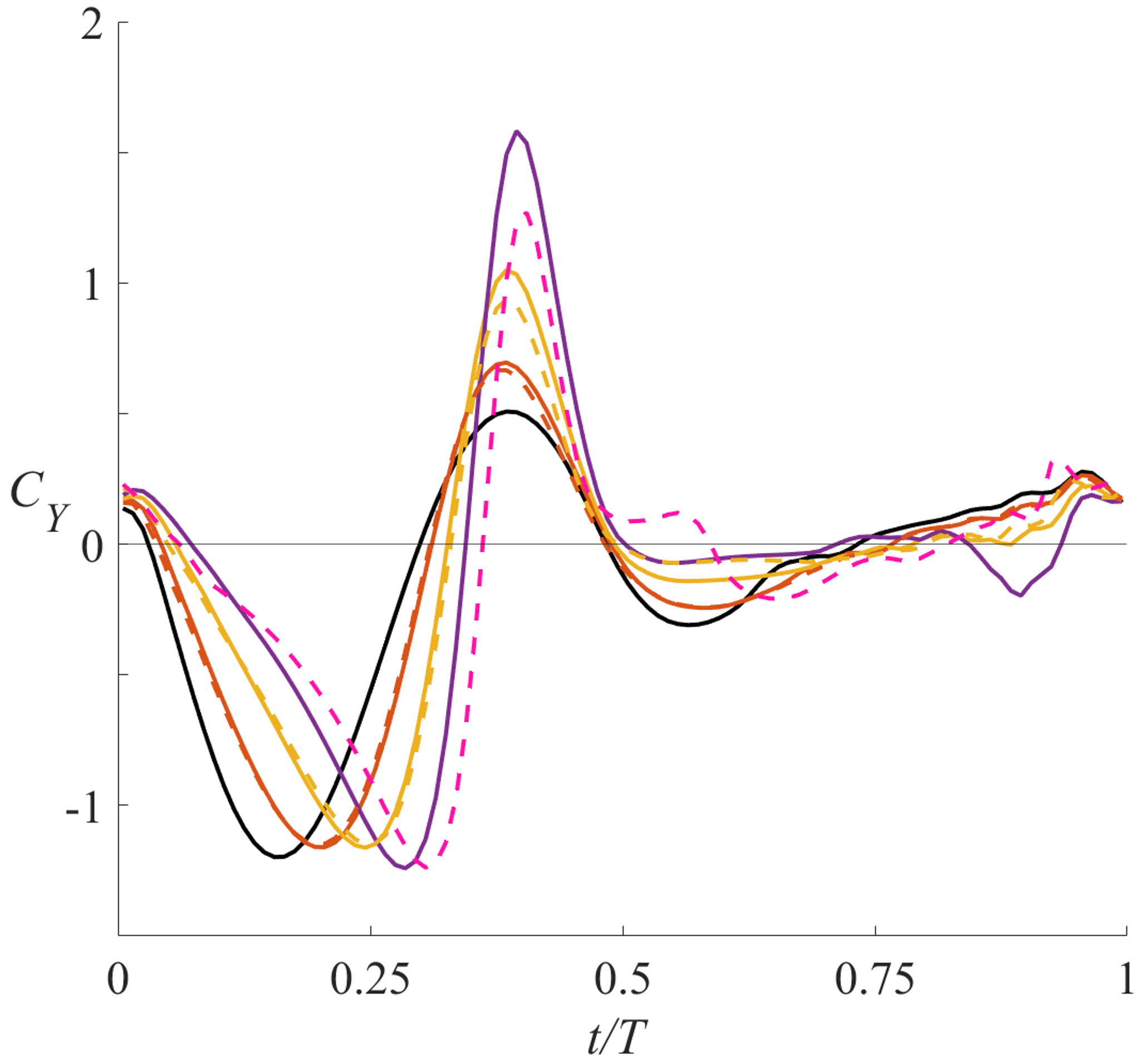}%
        \label{fig:3_2_b_forces}%
    \end{subfigure}
    \vspace{-0.25cm}%
    \caption{Streamwise (a) and cross-stream (b) forces on a single blade for different amplitudes at $\overline{\lambda} = 2 \text{ and } \lambda_\theta=2, \phi = 117^\circ $.}%
    \label{fig:tsr2000_forces}%
    \vspace{-0.25cm}%
\end{figure}

Implementing intracycle control also changes the streamwise and cross-stream blade-level forces, as shown in Figure \ref{fig:tsr2000_forces}. 
Despite the higher amplitude of control ($A_\lambda=0.83$), under a constant phase-averaged TSR, the turbine blades see lower peak streamwise and cross-stream forces than for time-averaged TSR at $A_\lambda=0.64$. Peak forces are also slightly lower under constant phase-averaged TSR than time-averaged at low-mid $A_\lambda$. 
At a constant time-averaged TSR, the peak magnitude of the cross-stream force (Fig. \ref{fig:3_2_b_forces}) is not affected significantly until the highest amplitude of intracycle control is implemented, where an increase of $32\%$ occurs when compared with the baseline. Peak streamwise force (Fig. \ref{fig:3_2_a_forces}) increases with the amplitude of control, ranging from $+19\%$ for $A_\lambda=0.18$ to $+96\%$ for $A_\lambda=0.64$, compared with the baseline. 

\vspace{-0.75cm}
\subsection{Operation in non-optimal TSR regimes}
\label{subsec:newRegimes}\vspace{-0.5cm}
Results in Section \ref{subsec:effTSR} find minimal power improvements over constant TSR using intracycle control at $\overline{\lambda}=2$, which is the optimal TSR of the turbine operating at constant angular velocity. However, results in Section \ref{subsec:validation} show that intracycle control may have more potential for improvement at lower time-averaged TSRs, or when $\overline{\lambda}<2$. 
Thus, this section explores intracycle simulations below ($ \overline{\lambda}<2 $) and above ($ \overline{\lambda}>2 $) the optimal constant speed TSR for three amplitudes ($A_\lambda=0.18,0.4,\text{ and }0.64$) and five phase shifts. 
Note that to keep the time-averaged TSR constant at $\overline{\lambda}=1.54$, the phase-averaged TSR, $\lambda_\theta$, varies from 1.56 to 1.82, and similarly for constant $ \overline{\lambda}=2.5 $, $\lambda_\theta$ varies from 2.55 to 2.97. 


\begin{figure}[htbp]
    \centering%
    \includegraphics[width=0.9\textwidth]{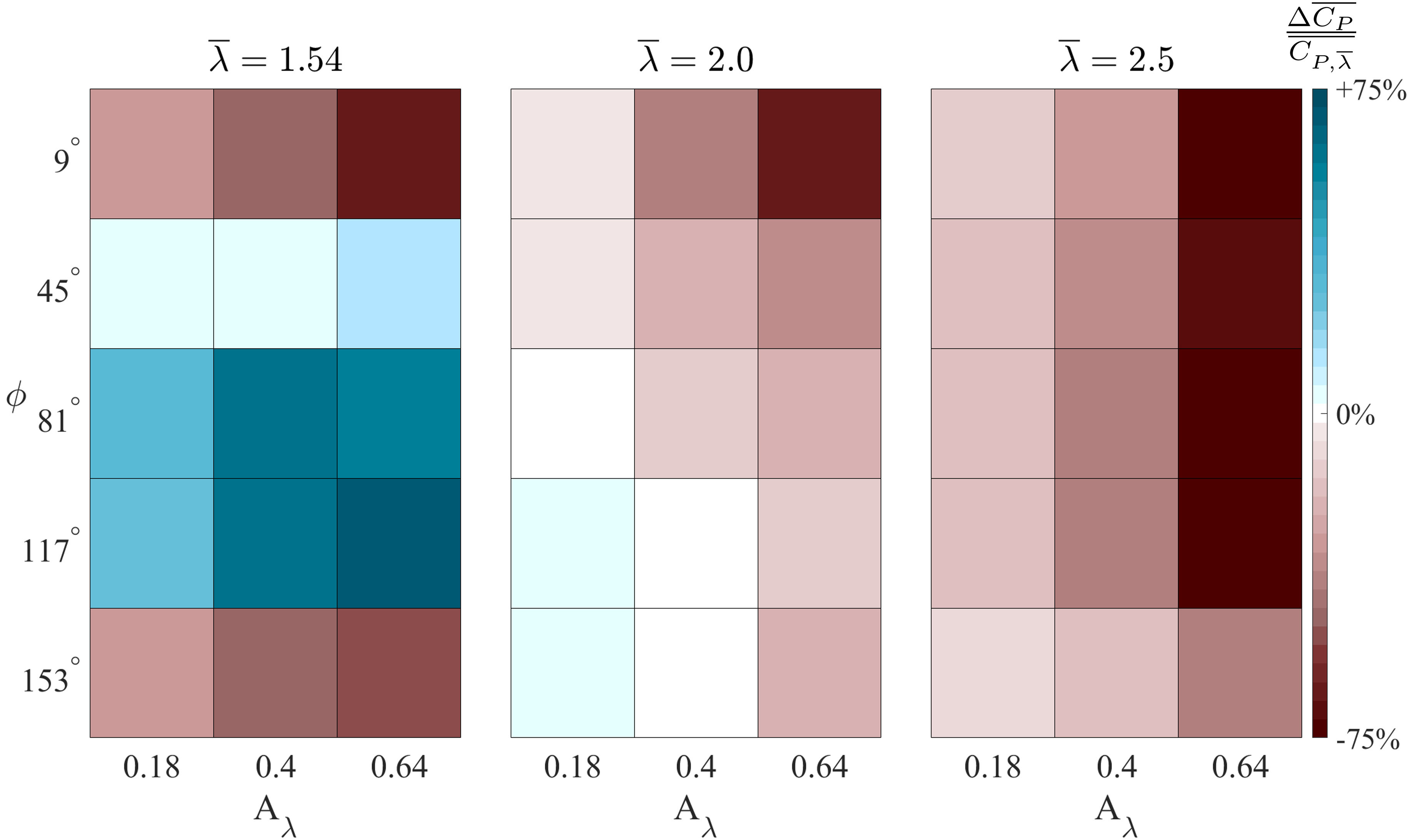}%
    \vspace{-0.25cm}
    \caption{Power difference with respect to constant TSR matching time-averaged intracycle TSR 
    baseline between intracycle control and constant rotation at three time-averaged TSRs: below-, at-, and above-optimal.}%
    \label{fig:phiContours}%
    \vspace{-0.25cm}
\end{figure}

Figure \ref{fig:phiContours} summarizes the changes in power between intracycle and constant TSR while holding time-averaged TSR constant. Contour levels indicate the percent change in power coefficient at each intracycle phase shift and amplitude. At above-optimal TSR ($ \overline{\lambda}=2.5 $), there is no improvement to be gained from intracycle velocity control, and power is decreased by $7\text{ - }78\%$ for the intracycle kinematics simulated. At optimal TSR, only the lowest amplitude cases improve power generation, and do so by nominal amounts of $5 \text{ - }6\%$, as discussed in Section \ref{subsec:effTSR}. Most cases at $\overline{\lambda}=2$ perform worse than the optimal constant speed case, reducing power by up to $64\%$. 

At a below-optimal TSR ($ \overline{\lambda}=1.54 $), however, a range of intracycle phase shifts are found that can perform better than the corresponding constant TSR for all intracycle amplitudes. There is a clear pattern with phase shift where alignment of the highest speed and/or acceleration with key phases of dynamic stall improves power generation. At $\phi=117 ^\circ $, the strongest trend across amplitudes is demonstrated, where improvement increases with amplitude, reaching $71\%$ improvement at $A_\lambda=0.64$ when compared with power at the equivalent constant speed. However, even for other phase shifts, there is still improvement, from $29\text{ - }60\%$ across $\phi=81^\circ$ and $3\text{ - }11\%$ across $\phi=45^\circ$. From $\phi=153\text{ - }9^\circ$, power is decreased by $29\text{ - }64\%$, where control slows the turbine blades during the power generation portion of the cycle and greatly increases flow separation from the blade. 

\begin{figure}[htbp]
    \centering%
    \includegraphics[width=1\textwidth]{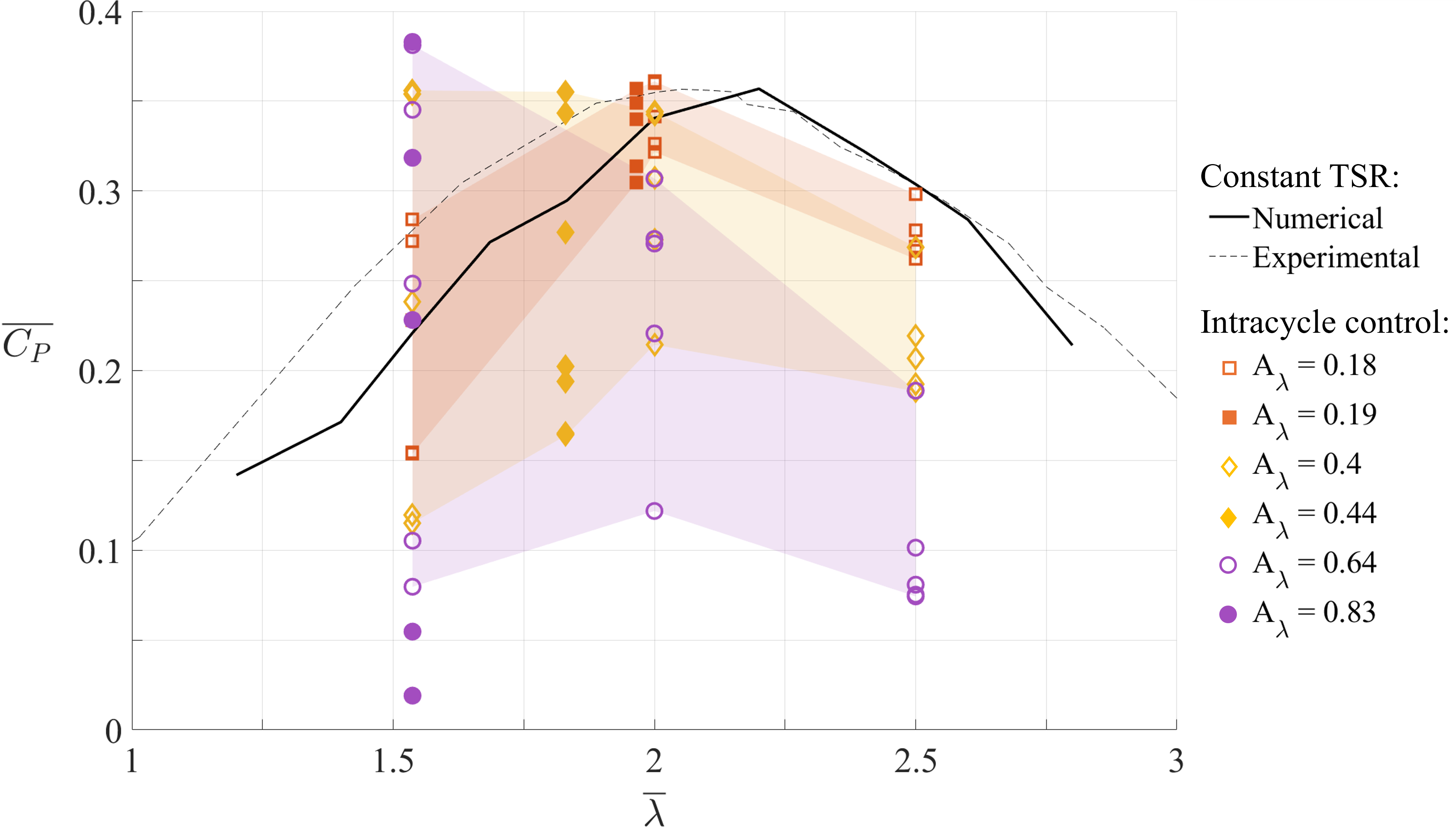}%
    \vspace{-0.25cm}
    \caption{Power coefficients for all intracycle control parameters implemented compared with constant TSR curve from simulations and experiments. Hollow markers correspond to numerical data at $\overline{\lambda}=1.54,2,\text{ and }2.5$ and $A_\lambda=0.18, 0.4$ and 0.64. Filled markers denote Table \ref{tab:expComp} computational data.}%
    \label{fig:tsrCurveWithIntracycle}%
    \vspace{-0.25cm}
\end{figure}

Visualized another way, Figure \ref{fig:tsrCurveWithIntracycle} shows the power coefficients for all simulations, including Table \ref{tab:expComp}, grouped by time-averaged TSR. Baseline conditions at constant angular velocity are shown for simulations and experiments as solid and dashed black lines, respectively. Markers denote the amplitude of intracycle velocity control, with orange squares at low-amplitude, yellow diamonds at medium-amplitude, and purple circles at high-amplitude. Hollow markers denote sets of the same low-, mid-, and high-amplitudes at $\overline{\lambda}=1.54,2,\text{ and }2.5$, and filled markers show Table \ref{tab:expComp} data, where phase-averaged TSR was held constant at $\lambda_\theta=2$. Shaded regions encompass all phase shifts for $A_\lambda=0.18, 0.4, \text{ and }0.64$. The highlighted regions also give insight as to how much improvement can be made in power generation with intracycle velocity control and the trends over different time-averaged TSRs and amplitudes, while not distinguishing between each phase shift. 

In general, the best performance (peak power) at a given TSR decreases as time-averaged TSR increases. At a time-averaged TSR lower than the constant speed optimum ($\overline{\lambda}=1.54$), the highest amplitude control produces both the upper and lower $ \overline{C_P} $ extremes, corresponding to opposing phase shift angles. The low-amplitude control exerts the smallest perturbations from the baseline constant TSR curve whereas the highest amplitude results in the the largest differences, reflected by the width of the corresponding shaded region. Conversely, in the regime above the optimal constant speed TSR ($\overline{\lambda}=2.5$), the shaded region stratifies between the low-middle-high amplitudes, where higher amplitudes produce the worst performance on average, and all control is detrimental. Thus, no improvement is found with intracycle control when time-averaged TSR is above optimal. 

\begin{figure}[htbp]
    \centering%
    \includegraphics[width=0.5\textwidth]{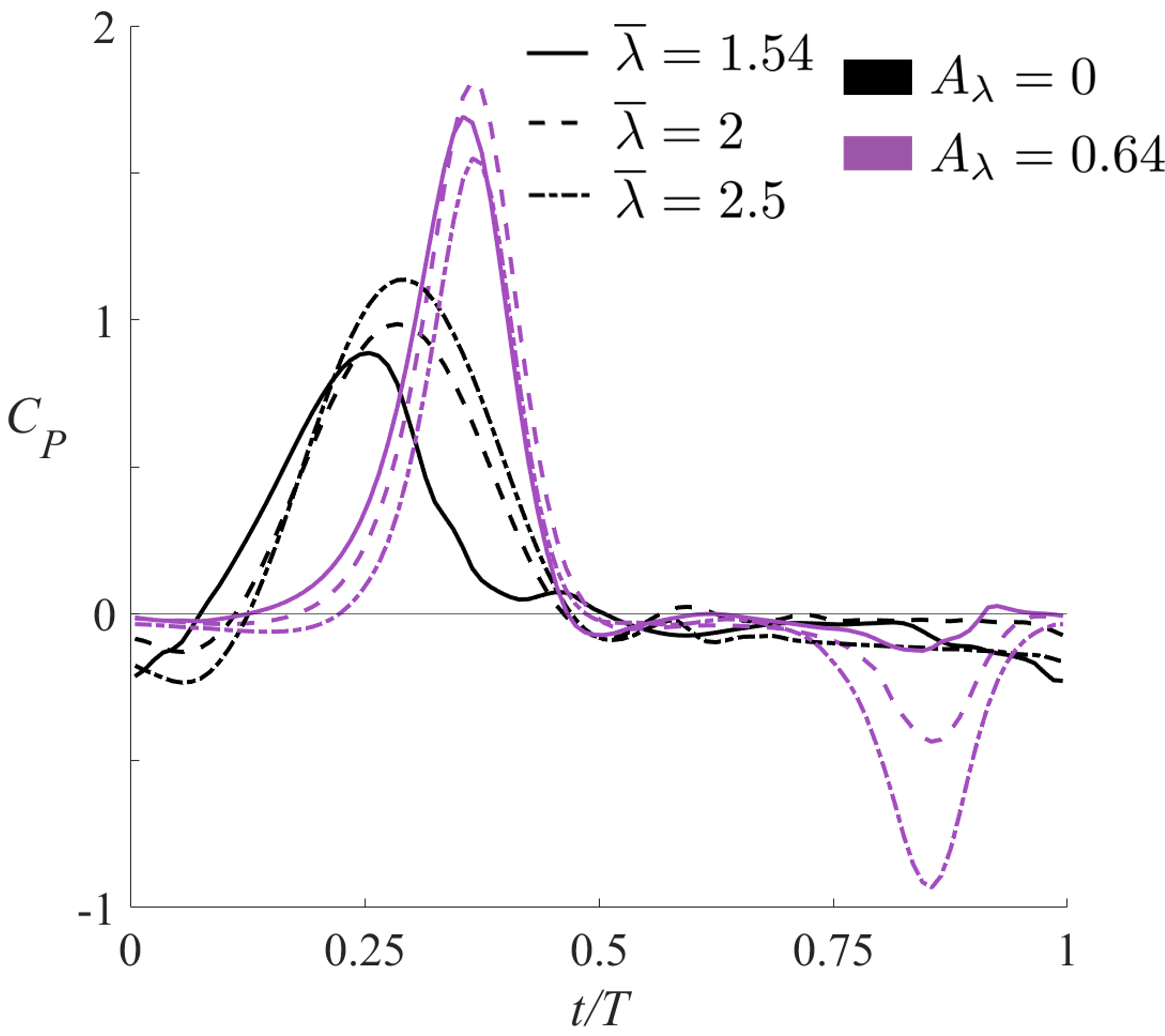}%
    \vspace{-0.25cm}
    \caption{Power vs time for constant and high-amplitude intracycle control at $\phi=117^\circ$ for three distinct time-averaged TSRs.}%
    \label{fig:aboveOpt_power}%
    \vspace{-0.5cm}
\end{figure}

\begin{figure}[htbp]
    \centering%
    \includegraphics[width=0.95\textwidth]{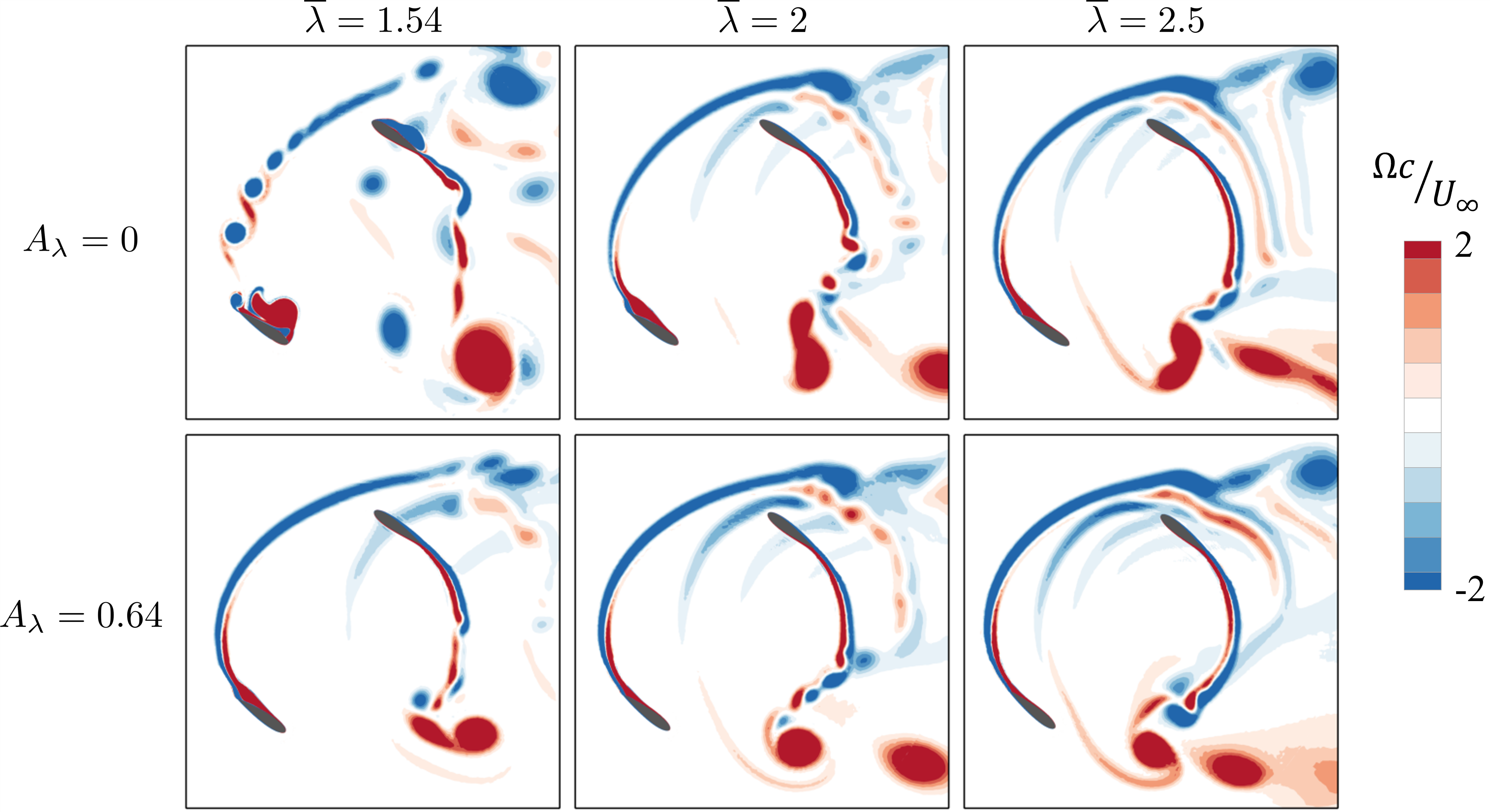}%
    \vspace{-0.25cm}
    \caption{Vorticity flow fields corresponding to power curves shown for constant and high-amplitude intracycle control at $\phi=117^\circ$.}%
    \label{fig:aboveOpt_vorticity}%
    \vspace{-0.25cm}
\end{figure}

Figure \ref{fig:aboveOpt_power} shows the power curve comparison between constant TSR and high-amplitude intracycle velocity control ($A_\lambda=0.64; \phi = 117^\circ $) in each TSR regime. Moving from below-optimal $ \overline{\lambda}=1.54 $ to above-optimal $ \overline{\lambda} = 2.5 $ reveals a decreasing peak power ($-8.4\%$) and power losses that are five times higher during the blade recovery stroke $(t/T>0.5)$. This corroborates findings by \citet{snortland2025downstream} that show blade performance during the recovery stroke to offset gains in power output at higher constant TSRs. Investigating the flow field in Figure \ref{fig:aboveOpt_vorticity} shows little change in the upstream flow attachment at $ \overline{\lambda} = 2.5 $ but stronger vortices in the wake of each blade that subsequent blades interact with as they move through the recovery stroke. This trend permeates all intracycle phase shifts, with the highest amplitudes of control performing the worst compared with constant TSR at $ \overline{\lambda} = 2.5 $, where power is decreased by $77.7\%$ (Fig. \ref{fig:phiContours}). Additionally, at a high amplitude of intracycle control, flow remains highly attached, with flow separation delayed through the end of power stroke for each time-averaged TSR, as shown in the bottom row of Figure \ref{fig:aboveOpt_vorticity}. 

\begin{figure}[htbp]
    \centering%
    \includegraphics[width=1\textwidth]{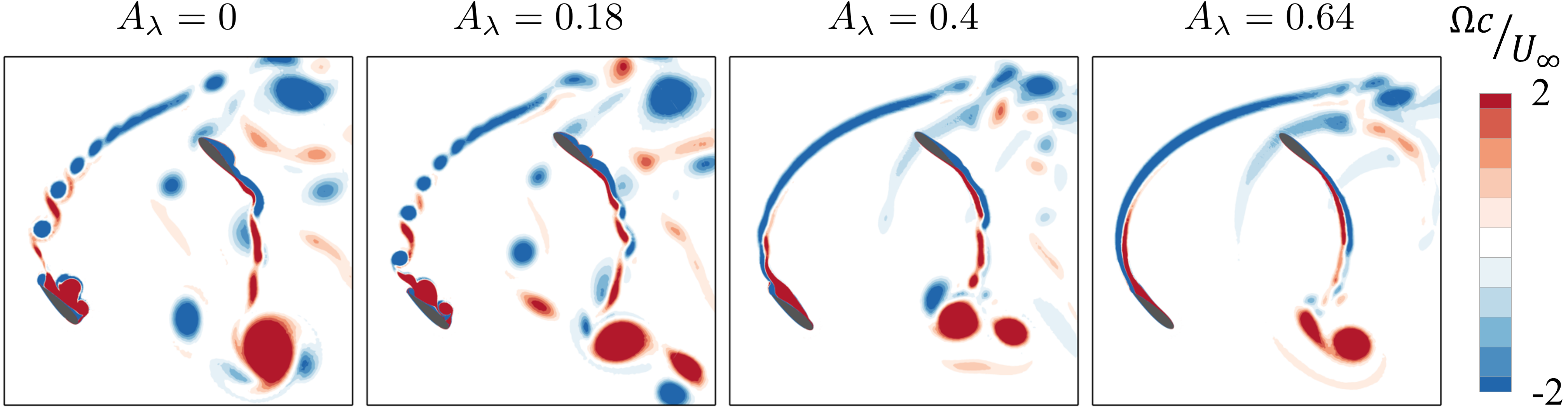}%
    \vspace{-0.25cm}
    \caption{Instantaneous spanwise vorticity flow fields at $ \overline{\lambda} = 1.54, \phi=117^\circ $ for each amplitude of intracycle velocity control.}%
    \label{fig:tsr_154_vorticity}%
    \vspace{-0.25cm}
\end{figure}

Broadly, results in Figures \ref{fig:phiContours} and \ref{fig:tsrCurveWithIntracycle} show that once the turbine is operating at optimal constant speed TSR, it is difficult to make further improvements in power generation using intracycle control. This is a similar finding as prior work on pitching optimization at optimal and high TSR \cite{lefouest2024}. However, for below-optimal TSR, intracycle control can be exploited to improve power generation and widen the range of TSRs for turbine operation. 
A plausible explanation for the different effects of TSR on control is that moving from $\overline{\lambda}=2$ to 1.54 generates a flow state where improvements can still be obtained. For example, in the baseline case at $\overline{\lambda}=1.54$, the dynamic stall vortex has a long time to develop and thus reaches its maximum size and sheds during the power stroke (Fig. \ref{fig:tsr_154_vorticity} left frame). By accelerating the foil through the upstream portion of the stroke, intracycle control keeps the boundary layer attached longer, delays the dynamic stall vortex formation, and improves lift and power generation (Fig. \ref{fig:tsr_154_vorticity} right frame). 
Additionally, a similar trend is seen for the second blade as it finishes its recovery stroke, showing more attached flow and thus, less vortex shedding. The results improve the $\overline{C_P}$ not only compared to the constant TSR at $\overline{\lambda}=1.54$, but also exceed those of the constant TSR at $\overline{\lambda}=2$ by $4.5\text{ - }12.2\%$ for mid-high $A_\lambda$.  
One contributing factor to this improvement could be that operation at this lower TSR reduces the impact of detrimental downstream (recovery stroke) performance \cite{snortland2025downstream}. 

\begin{figure}[htbp]
    \centering%
    \includegraphics[width=0.5\textwidth]{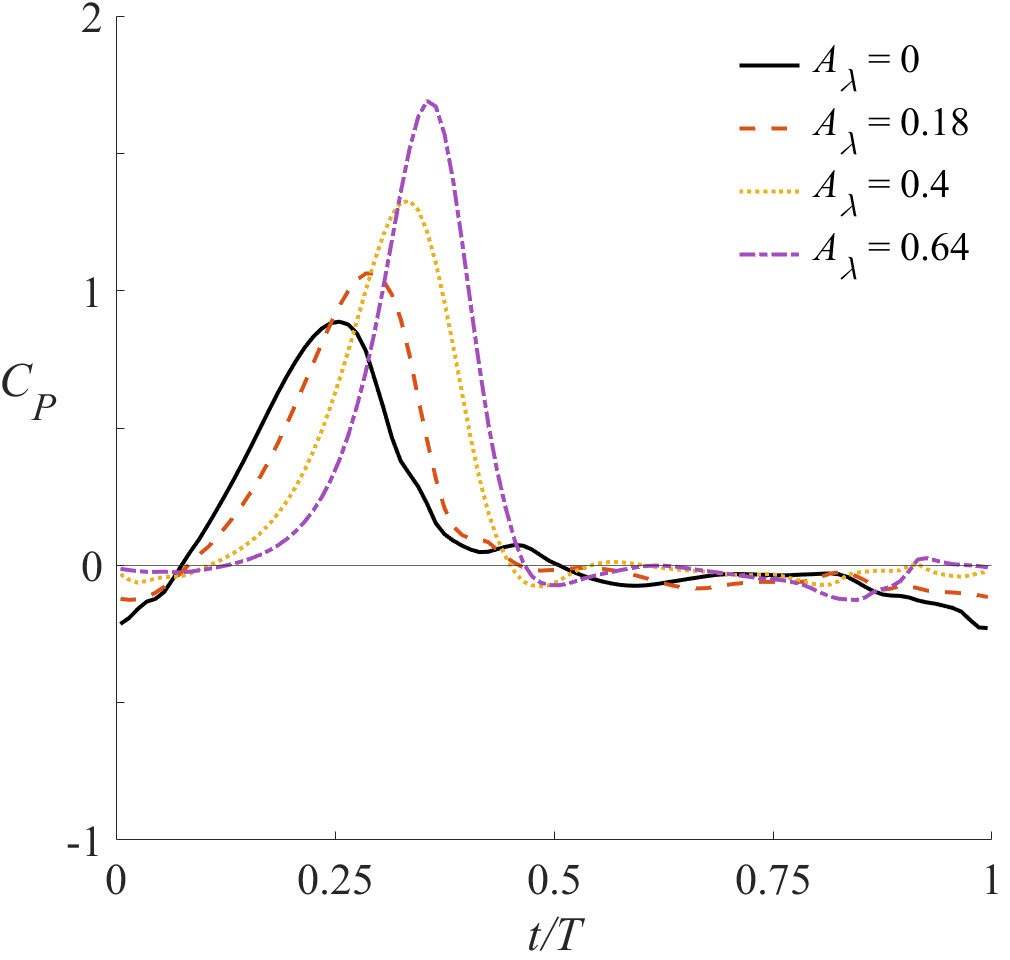}%
    \vspace{-0.25cm}
    \caption{Power vs time at $ \overline{\lambda} = 1.54, \phi=117^\circ $ for each amplitude of intracycle velocity control.}%
    \label{fig:tsr_154_power}%
    \vspace{-0.25cm}
\end{figure}

Looking more closely at $ \overline{\lambda} = 1.54 $, Figure \ref{fig:tsr_154_power} shows the phase-and-blade-averaged power curves over time for a constant TSR and under intracycle control at three amplitudes. Peak power is significantly higher for the high-amplitude intracycle control case, and loss of power during the second half of the cycle is comparable across all cases, consistent with what is often observed for constant speed at lower TSR, and considerably different to what is observed for $ \overline{\lambda} = 2$ in Figure \ref{fig:constEff_powerCurve}. 

\begin{figure}[htbp]
    \vspace{-0.25cm}
    \centering%
    \begin{subfigure}{0.48\textwidth}
        \subcaption{}\vspace{-0.25cm}%
        \includegraphics[width=1\textwidth]{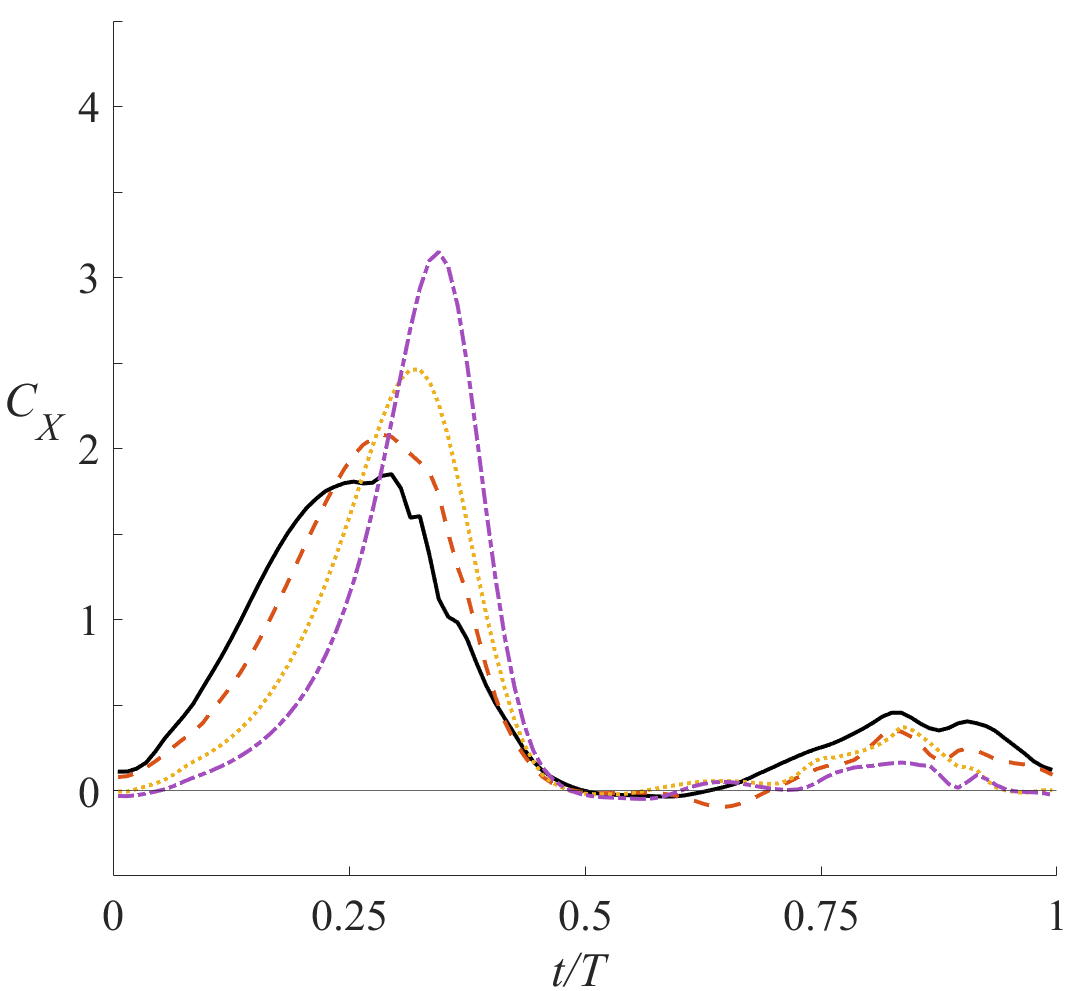}%
        \label{fig:tsr1537_forces_cx}%
    \end{subfigure}
    \begin{subfigure}{0.48\textwidth}%
        \subcaption{}\vspace{-0.25cm}%
        \includegraphics[width=1\textwidth]{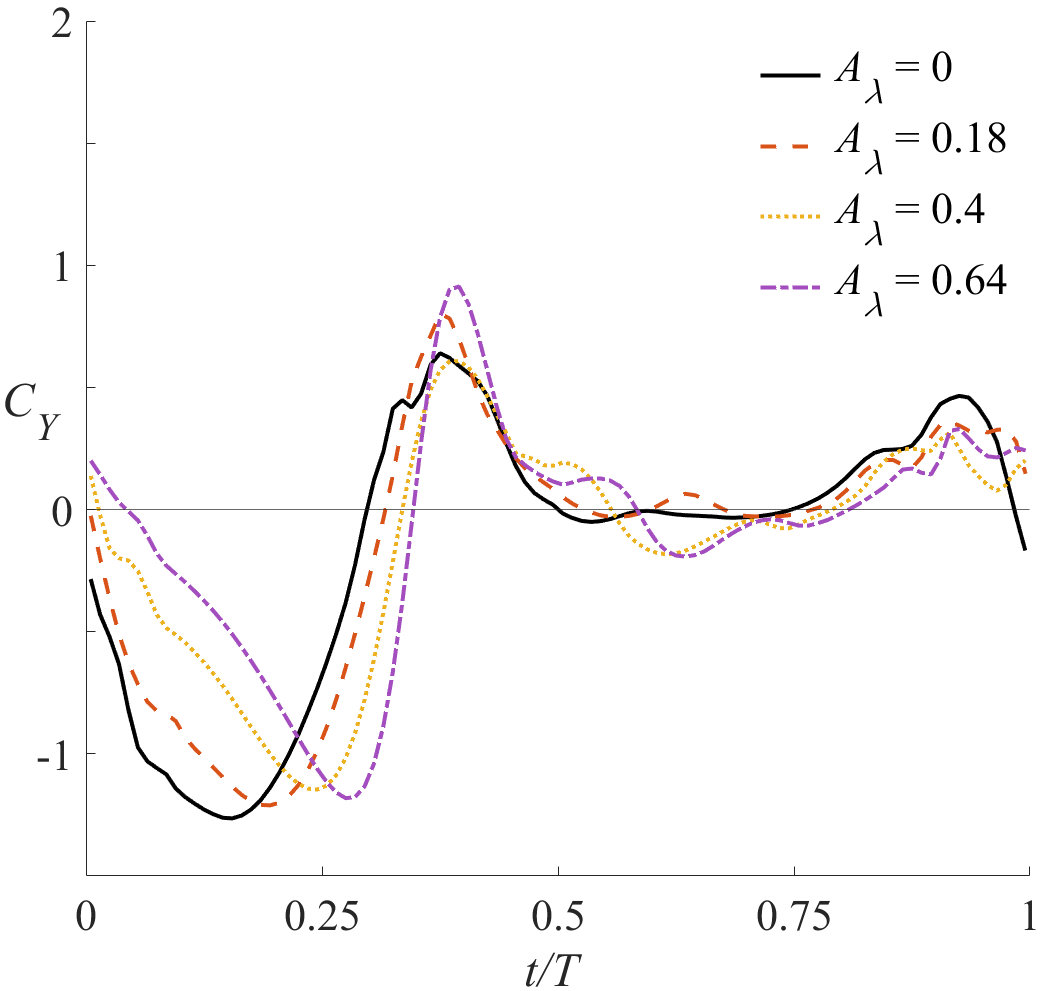}%
        \label{fig:tsr1537_forces_cy}%
    \end{subfigure}
    \vspace{-0.25cm}
    \caption{Streamwise (a) and cross-stream (b) forces on a single blade at $\overline{\lambda} = 1.54, \phi = 117^\circ $ for each amplitude of intracycle velocity control.}%
    \label{fig:tsr1537_forces}%
    \vspace{-0.5cm}
\end{figure}

Increases in overall power output from cross flow turbines are often associated with increases to loading and so Figure \ref{fig:tsr1537_forces} explores if this is also the case for intracycle control at $\overline{\lambda} = 1.54$. 
Figure \ref{fig:tsr1537_forces_cy} indicates that cross-stream forces are similar and sometimes favorably altered by intracycle control at this TSR when compared to constant speed at the same TSR, showing peak $|C_Y|$ decreases by $4\text{ - }9\%$. In contrast, Fig. \ref{fig:tsr1537_forces_cx} demonstrates an increase in peak streamwise forces due to intracycle control, with peak $C_X$ increasing by $12\%, 33\%, \text{ and } 70\%$ for the low, mid, and high amplitude values, respectively. The same amplitudes produce $27\%, 59\%, \text{ and } 71\%$ more power, correspondingly, and so the increase in peak loading is comparable to the increase in power output. 

\vspace{-0.5cm}
\section{Conclusions}
\label{sec:concl}\vspace{-0.5cm}

The effects of intracycle angular velocity control on a 2-bladed cross-flow turbine performance, forces, and flow dynamics are investigated at laboratory scale for a chord-based Reynolds number of $Re_{\text{c}}=45,000$. The aim is to explore differences in the effects of intracycle control over a range of TSRs and to identify regimes where overall improvement in performance can be obtained. RANS simulations are validated with experimental results at $\lambda_\theta=2$, very close to the optimal TSR for this turbine geometry when operated under constant speed. A sweep of intracycle phase shifts and amplitude of sinusoidal oscillations is applied at a phase-averaged TSR of $\lambda_\theta=2$ and show simulation results that match well with experimental data in both power metrics and flow fields. 

It is demonstrated that, when intracycle control is applied, the time-averaged TSR is lowered due to the nonlinear relationship between blade position and time. This result is amplified with increasing $A_{\lambda}$. This indicates that performance differences resulting from intracycle control could be caused by two competing factors: 1) a change in mean TSR with respect to time, and 2) the local acceleration and deceleration of the blade. Thus, a time-averaged TSR, $\overline{\lambda}$, is utilized in order to systematically assess the control parameter space, which is comprised of an amplitude $A_\lambda$ and phase shift $\phi$. 

Intracycle simulations are performed at {\it time-averaged} TSRs of $\overline{\lambda}=$ 1.54, $2$, and $2.5$, by varying the amplitude and phase shift of control. When compared with the baseline (constant angular velocity at equivalent TSR), the largest benefits of intracycle control are found below the optimal constant TSR. At $\overline{\lambda}=1.54$,  improvements in power of up to $70.5\%$ are reported along with comparable increase in the peak blade forces. At $A_\lambda=0.64$ and $\phi=117^\circ$, the power coefficient of $\overline{C_{\text{P}}}=0.377$ is $12.2\%$ higher than the peak $\overline{C_{\text{P}}}$ under constant velocity at $\overline{\lambda}=2$. 
Improvements are achieved by shifting the time-averaged TSR to a lower regime where there is more potential to improve performance during the power stroke and less detrimental impact during the recovery stroke. The vortex that forms on the blade sheds earlier in the power stroke as it has more time to develop, and beneficial intracycle control kinematics promote boundary layer attachment to the blades through the acceleration portion of the control and thus delay separation until the power stroke is near complete. In contrast, at phase shifts when intracycle control is detrimental to power, peak forces are exacerbated by decelerating the blade through vortex roll up and triggering even early flow separation from the blades. 

For other TSR regimes explored, intracycle control is not found to be as effective. Control does not improve power generation at $\overline{\lambda}=2.5$, and produces only minimal improvements of up to $6.7\%$ at $\overline{\lambda}=2$. Low amplitude control can improve power generation at optimal TSR, but the tradeoffs in peak forces may offset these gains. The lack of significant improvements in power generation at $\overline{\lambda}=2$ and $2.5$ seem to be because, at these TSRs, the flow is already fully attached for the majority of the power stroke; thus, additional acceleration via intracycle control yields little change since it cannot further exploit those mechanics to improve performance. Additionally, performance during the blade recovery stroke is detrimental for above-optimal TSRs and offsets any benefits seen during the power stroke.
\vfill
\begin{acknowledgments}\vspace{-0.5cm}
This work was supported by the National Science Foundation Graduate Research Fellowship Program (SFH: Grant No. 2137424), the TEAMER program under Department of Energy (JAF/OW: Project number EE0008895), and U.S. Department of Defense's Naval Facilities Engineering Systems Command (AA: N0002421D6400/N0002421F8712).
\end{acknowledgments}

\section*{Author Declarations}\vspace{-0.5cm}
\subsection*{Author Contributions}\vspace{-0.5cm}
Conceptualization and Funding: JAF and OW. Methodology: All authors. Investigation and Analysis: SFH and AA. Writing - original draft: SFH. Writing - review and editing: All authors.
\vspace{-0.5cm}
\subsection*{Conflicts of interest}\vspace{-0.5cm}
The authors have no conflicts to disclose. 

\section*{Data Availability Statement}\vspace{-0.5cm}
The data from section \ref{subsec:validation} is available online via the Marine and Hydrokinetic Data Repository (MHKDR). The data from sections \ref{subsec:effTSR} and \ref{subsec:newRegimes} is available from the corresponding author upon reasonable request. 

\bibliography{cft_literature}

\begin{thebibliography}{30}%
\makeatletter
\providecommand \@ifxundefined [1]{%
 \@ifx{#1\undefined}
}%
\providecommand \@ifnum [1]{%
 \ifnum #1\expandafter \@firstoftwo
 \else \expandafter \@secondoftwo
 \fi
}%
\providecommand \@ifx [1]{%
 \ifx #1\expandafter \@firstoftwo
 \else \expandafter \@secondoftwo
 \fi
}%
\providecommand \natexlab [1]{#1}%
\providecommand \enquote  [1]{``#1''}%
\providecommand \bibnamefont  [1]{#1}%
\providecommand \bibfnamefont [1]{#1}%
\providecommand \citenamefont [1]{#1}%
\providecommand \href@noop [0]{\@secondoftwo}%
\providecommand \href [0]{\begingroup \@sanitize@url \@href}%
\providecommand \@href[1]{\@@startlink{#1}\@@href}%
\providecommand \@@href[1]{\endgroup#1\@@endlink}%
\providecommand \@sanitize@url [0]{\catcode `\\12\catcode `\$12\catcode `\&12\catcode `\#12\catcode `\^12\catcode `\_12\catcode `\%12\relax}%
\providecommand \@@startlink[1]{}%
\providecommand \@@endlink[0]{}%
\providecommand \url  [0]{\begingroup\@sanitize@url \@url }%
\providecommand \@url [1]{\endgroup\@href {#1}{\urlprefix }}%
\providecommand \urlprefix  [0]{URL }%
\providecommand \Eprint [0]{\href }%
\providecommand \doibase [0]{http://dx.doi.org/}%
\providecommand \selectlanguage [0]{\@gobble}%
\providecommand \bibinfo  [0]{\@secondoftwo}%
\providecommand \bibfield  [0]{\@secondoftwo}%
\providecommand \translation [1]{[#1]}%
\providecommand \BibitemOpen [0]{}%
\providecommand \bibitemStop [0]{}%
\providecommand \bibitemNoStop [0]{.\EOS\space}%
\providecommand \EOS [0]{\spacefactor3000\relax}%
\providecommand \BibitemShut  [1]{\csname bibitem#1\endcsname}%
\let\auto@bib@innerbib\@empty
\bibitem [{\citenamefont {Rezaeiha}, \citenamefont {Montazeri},\ and\ \citenamefont {Blocken}(2018)}]{rezaeiha2018}%
  \BibitemOpen
  \bibfield  {author} {\bibinfo {author} {\bibfnamefont {A.}~\bibnamefont {Rezaeiha}}, \bibinfo {author} {\bibfnamefont {H.}~\bibnamefont {Montazeri}}, \ and\ \bibinfo {author} {\bibfnamefont {B.}~\bibnamefont {Blocken}},\ }\bibfield  {title} {\enquote {\bibinfo {title} {Towards optimal aerodynamic design of vertical axis wind turbines: {Impact} of solidity and number of blades},}\ }\href {\doibase 10.1016/j.energy.2018.09.192} {\bibfield  {journal} {\bibinfo  {journal} {Energy}\ }\textbf {\bibinfo {volume} {165}},\ \bibinfo {pages} {1129--1148} (\bibinfo {year} {2018})}\BibitemShut {NoStop}%
\bibitem [{\citenamefont {Duarte}\ \emph {et~al.}(2022)\citenamefont {Duarte}, \citenamefont {Maguin}, \citenamefont {Dellinger}, \citenamefont {Dellinger},\ and\ \citenamefont {Vazquez}}]{duarte2022}%
  \BibitemOpen
  \bibfield  {author} {\bibinfo {author} {\bibfnamefont {L.}~\bibnamefont {Duarte}}, \bibinfo {author} {\bibfnamefont {N.}~\bibnamefont {Maguin}}, \bibinfo {author} {\bibfnamefont {G.}~\bibnamefont {Dellinger}}, \bibinfo {author} {\bibfnamefont {N.}~\bibnamefont {Dellinger}}, \ and\ \bibinfo {author} {\bibfnamefont {J.}~\bibnamefont {Vazquez}},\ }\bibfield  {title} {\enquote {\bibinfo {title} {Numerical investigation of a two-bladed vertical-axis turbine operating in a confined channel},}\ }\href {\doibase 10.1016/j.ecmx.2022.100298} {\bibfield  {journal} {\bibinfo  {journal} {Energy Conversion and Management: X}\ }\textbf {\bibinfo {volume} {16}},\ \bibinfo {pages} {100298} (\bibinfo {year} {2022})}\BibitemShut {NoStop}%
\bibitem [{\citenamefont {Kumar}\ and\ \citenamefont {Sarkar}(2022)}]{kumar2022}%
  \BibitemOpen
  \bibfield  {author} {\bibinfo {author} {\bibfnamefont {R.}~\bibnamefont {Kumar}}\ and\ \bibinfo {author} {\bibfnamefont {S.}~\bibnamefont {Sarkar}},\ }\bibfield  {title} {\enquote {\bibinfo {title} {Effect of design parameters on the performance of helical {Darrieus} hydrokinetic turbines},}\ }\href {\doibase 10.1016/j.rser.2022.112431} {\bibfield  {journal} {\bibinfo  {journal} {Renewable and Sustainable Energy Reviews}\ }\textbf {\bibinfo {volume} {162}},\ \bibinfo {pages} {112431} (\bibinfo {year} {2022})}\BibitemShut {NoStop}%
\bibitem [{\citenamefont {Araya}, \citenamefont {Colonius},\ and\ \citenamefont {Dabiri}(2017)}]{araya2017}%
  \BibitemOpen
  \bibfield  {author} {\bibinfo {author} {\bibfnamefont {D.~B.}\ \bibnamefont {Araya}}, \bibinfo {author} {\bibfnamefont {T.}~\bibnamefont {Colonius}}, \ and\ \bibinfo {author} {\bibfnamefont {J.~O.}\ \bibnamefont {Dabiri}},\ }\bibfield  {title} {\enquote {\bibinfo {title} {Transition to bluff-body dynamics in the wake of vertical-axis wind turbines},}\ }\href {\doibase 10.1017/jfm.2016.862} {\bibfield  {journal} {\bibinfo  {journal} {Journal of Fluid Mechanics}\ }\textbf {\bibinfo {volume} {813}},\ \bibinfo {pages} {346--381} (\bibinfo {year} {2017})}\BibitemShut {NoStop}%
\bibitem [{\citenamefont {Hunt}\ \emph {et~al.}(2024)\citenamefont {Hunt}, \citenamefont {Strom}, \citenamefont {Talpey}, \citenamefont {Ross}, \citenamefont {Scherl}, \citenamefont {Brunton}, \citenamefont {Wosnik},\ and\ \citenamefont {Polagye}}]{hunt2024}%
  \BibitemOpen
  \bibfield  {author} {\bibinfo {author} {\bibfnamefont {A.}~\bibnamefont {Hunt}}, \bibinfo {author} {\bibfnamefont {B.}~\bibnamefont {Strom}}, \bibinfo {author} {\bibfnamefont {G.}~\bibnamefont {Talpey}}, \bibinfo {author} {\bibfnamefont {H.}~\bibnamefont {Ross}}, \bibinfo {author} {\bibfnamefont {I.}~\bibnamefont {Scherl}}, \bibinfo {author} {\bibfnamefont {S.}~\bibnamefont {Brunton}}, \bibinfo {author} {\bibfnamefont {M.}~\bibnamefont {Wosnik}}, \ and\ \bibinfo {author} {\bibfnamefont {B.}~\bibnamefont {Polagye}},\ }\bibfield  {title} {\enquote {\bibinfo {title} {An experimental evaluation of the interplay between geometry and scale on cross-flow turbine performance},}\ }\href {\doibase 10.1016/j.rser.2024.114848} {\bibfield  {journal} {\bibinfo  {journal} {Renewable and Sustainable Energy Reviews}\ }\textbf {\bibinfo {volume} {206}},\ \bibinfo {pages} {114848} (\bibinfo {year} {2024})}\BibitemShut {NoStop}%
\bibitem [{\citenamefont {Snortland}, \citenamefont {Polagye},\ and\ \citenamefont {Williams}(2019)}]{snortland2019}%
  \BibitemOpen
  \bibfield  {author} {\bibinfo {author} {\bibfnamefont {A.}~\bibnamefont {Snortland}}, \bibinfo {author} {\bibfnamefont {B.}~\bibnamefont {Polagye}}, \ and\ \bibinfo {author} {\bibfnamefont {O.}~\bibnamefont {Williams}},\ }\bibfield  {title} {\enquote {\bibinfo {title} {Influence of near-blade hydrodynamics on cross-flow turbine performance},}\ }\href@noop {} {\bibfield  {journal} {\bibinfo  {journal} {Proceedings of the European Wave and Tidal Energy Conference}\ }\textbf {\bibinfo {volume} {13}} (\bibinfo {year} {2019})}\BibitemShut {NoStop}%
\bibitem [{\citenamefont {Dave}\ and\ \citenamefont {Franck}(2023)}]{dave2023}%
  \BibitemOpen
  \bibfield  {author} {\bibinfo {author} {\bibfnamefont {M.}~\bibnamefont {Dave}}\ and\ \bibinfo {author} {\bibfnamefont {J.~A.}\ \bibnamefont {Franck}},\ }\bibfield  {title} {\enquote {\bibinfo {title} {Analysis of dynamic stall development on a cross-flow turbine blade},}\ }\href {\doibase 10.1103/PhysRevFluids.8.074702} {\bibfield  {journal} {\bibinfo  {journal} {Physical Review Fluids}\ }\textbf {\bibinfo {volume} {8}},\ \bibinfo {pages} {074702} (\bibinfo {year} {2023})}\BibitemShut {NoStop}%
\bibitem [{\citenamefont {Le~Fouest}\ and\ \citenamefont {Mulleners}(2022)}]{lefouest2022}%
  \BibitemOpen
  \bibfield  {author} {\bibinfo {author} {\bibfnamefont {S.}~\bibnamefont {Le~Fouest}}\ and\ \bibinfo {author} {\bibfnamefont {K.}~\bibnamefont {Mulleners}},\ }\bibfield  {title} {\enquote {\bibinfo {title} {The dynamic stall dilemma for vertical-axis wind turbines},}\ }\href {\doibase 10.1016/j.renene.2022.07.071} {\bibfield  {journal} {\bibinfo  {journal} {Renewable Energy}\ }\textbf {\bibinfo {volume} {198}},\ \bibinfo {pages} {505--520} (\bibinfo {year} {2022})}\BibitemShut {NoStop}%
\bibitem [{\citenamefont {Allet}, \citenamefont {Brahimi},\ and\ \citenamefont {Paraschivoiu}(1997)}]{allet1997}%
  \BibitemOpen
  \bibfield  {author} {\bibinfo {author} {\bibfnamefont {A.}~\bibnamefont {Allet}}, \bibinfo {author} {\bibfnamefont {M.}~\bibnamefont {Brahimi}}, \ and\ \bibinfo {author} {\bibfnamefont {I.}~\bibnamefont {Paraschivoiu}},\ }\bibfield  {title} {\enquote {\bibinfo {title} {On the {Aerodynamic} modeling of a {VAWT}},}\ }\href {https://www.jstor.org/stable/43749657} {\bibfield  {journal} {\bibinfo  {journal} {Wind Engineering}\ }\textbf {\bibinfo {volume} {21}},\ \bibinfo {pages} {351--365} (\bibinfo {year} {1997})}\BibitemShut {NoStop}%
\bibitem [{\citenamefont {Mandal}\ and\ \citenamefont {Burton}(1994)}]{mandal1994}%
  \BibitemOpen
  \bibfield  {author} {\bibinfo {author} {\bibfnamefont {A.}~\bibnamefont {Mandal}}\ and\ \bibinfo {author} {\bibfnamefont {J.}~\bibnamefont {Burton}},\ }\bibfield  {title} {\enquote {\bibinfo {title} {The {Effects} of {Dynamic} {Stall} and {Flow} {Curvature} on the {Aerodynamics} of {Darrieus} {Turbines} {Applying} the {Cascade} {Model}},}\ }\href {https://www.jstor.org/stable/43749553} {\bibfield  {journal} {\bibinfo  {journal} {Wind Engineering}\ }\textbf {\bibinfo {volume} {18}},\ \bibinfo {pages} {267--282} (\bibinfo {year} {1994})}\BibitemShut {NoStop}%
\bibitem [{\citenamefont {Simão~Ferreira}\ \emph {et~al.}(2009)\citenamefont {Simão~Ferreira}, \citenamefont {van Kuik}, \citenamefont {van Bussel},\ and\ \citenamefont {Scarano}}]{ferreira2009}%
  \BibitemOpen
  \bibfield  {author} {\bibinfo {author} {\bibfnamefont {C.}~\bibnamefont {Simão~Ferreira}}, \bibinfo {author} {\bibfnamefont {G.}~\bibnamefont {van Kuik}}, \bibinfo {author} {\bibfnamefont {G.}~\bibnamefont {van Bussel}}, \ and\ \bibinfo {author} {\bibfnamefont {F.}~\bibnamefont {Scarano}},\ }\bibfield  {title} {\enquote {\bibinfo {title} {Visualization by {PIV} of dynamic stall on a vertical axis wind turbine},}\ }\href {\doibase 10.1007/s00348-008-0543-z} {\bibfield  {journal} {\bibinfo  {journal} {Experiments in Fluids}\ }\textbf {\bibinfo {volume} {46}},\ \bibinfo {pages} {97--108} (\bibinfo {year} {2009})}\BibitemShut {NoStop}%
\bibitem [{\citenamefont {Amet}\ \emph {et~al.}(2009)\citenamefont {Amet}, \citenamefont {Maître}, \citenamefont {Pellone},\ and\ \citenamefont {Achard}}]{amet2009}%
  \BibitemOpen
  \bibfield  {author} {\bibinfo {author} {\bibfnamefont {E.}~\bibnamefont {Amet}}, \bibinfo {author} {\bibfnamefont {T.}~\bibnamefont {Maître}}, \bibinfo {author} {\bibfnamefont {C.}~\bibnamefont {Pellone}}, \ and\ \bibinfo {author} {\bibfnamefont {J.-L.}\ \bibnamefont {Achard}},\ }\bibfield  {title} {\enquote {\bibinfo {title} {{2D} {Numerical} simulations of blade-vortex interaction in a darrieus turbine},}\ }\href {\doibase 10.1115/1.4000258} {\bibfield  {journal} {\bibinfo  {journal} {Journal of Fluids Engineering}\ }\textbf {\bibinfo {volume} {131}},\ \bibinfo {pages} {111103} (\bibinfo {year} {2009})}\BibitemShut {NoStop}%
\bibitem [{\citenamefont {Tsang}\ \emph {et~al.}(2008)\citenamefont {Tsang}, \citenamefont {So}, \citenamefont {Leung},\ and\ \citenamefont {Wang}}]{tsang2008}%
  \BibitemOpen
  \bibfield  {author} {\bibinfo {author} {\bibfnamefont {K.~K.~Y.}\ \bibnamefont {Tsang}}, \bibinfo {author} {\bibfnamefont {R.~M.~C.}\ \bibnamefont {So}}, \bibinfo {author} {\bibfnamefont {R.~C.~K.}\ \bibnamefont {Leung}}, \ and\ \bibinfo {author} {\bibfnamefont {X.~Q.}\ \bibnamefont {Wang}},\ }\bibfield  {title} {\enquote {\bibinfo {title} {Dynamic stall behavior from unsteady force measurements},}\ }\href {\doibase 10.1016/j.jfluidstructs.2007.06.007} {\bibfield  {journal} {\bibinfo  {journal} {Journal of Fluids and Structures}\ }\textbf {\bibinfo {volume} {24}},\ \bibinfo {pages} {129--150} (\bibinfo {year} {2008})}\BibitemShut {NoStop}%
\bibitem [{\citenamefont {Zhao}\ \emph {et~al.}(2022)\citenamefont {Zhao}, \citenamefont {Wang}, \citenamefont {Wang}, \citenamefont {Shen}, \citenamefont {Liu},\ and\ \citenamefont {Chen}}]{zhao2022}%
  \BibitemOpen
  \bibfield  {author} {\bibinfo {author} {\bibfnamefont {Z.}~\bibnamefont {Zhao}}, \bibinfo {author} {\bibfnamefont {D.}~\bibnamefont {Wang}}, \bibinfo {author} {\bibfnamefont {T.}~\bibnamefont {Wang}}, \bibinfo {author} {\bibfnamefont {W.}~\bibnamefont {Shen}}, \bibinfo {author} {\bibfnamefont {H.}~\bibnamefont {Liu}}, \ and\ \bibinfo {author} {\bibfnamefont {M.}~\bibnamefont {Chen}},\ }\bibfield  {title} {\enquote {\bibinfo {title} {A review: {Approaches} for aerodynamic performance improvement of lift-type vertical axis wind turbine},}\ }\href {\doibase 10.1016/j.seta.2021.101789} {\bibfield  {journal} {\bibinfo  {journal} {Sustainable Energy Technologies and Assessments}\ }\textbf {\bibinfo {volume} {49}},\ \bibinfo {pages} {101789} (\bibinfo {year} {2022})}\BibitemShut {NoStop}%
\bibitem [{\citenamefont {Choudhry}, \citenamefont {Arjomandi},\ and\ \citenamefont {Kelso}(2016)}]{choudhry2016}%
  \BibitemOpen
  \bibfield  {author} {\bibinfo {author} {\bibfnamefont {A.}~\bibnamefont {Choudhry}}, \bibinfo {author} {\bibfnamefont {M.}~\bibnamefont {Arjomandi}}, \ and\ \bibinfo {author} {\bibfnamefont {R.}~\bibnamefont {Kelso}},\ }\bibfield  {title} {\enquote {\bibinfo {title} {Methods to control dynamic stall for wind turbine applications},}\ }\href {\doibase 10.1016/j.renene.2015.07.097} {\bibfield  {journal} {\bibinfo  {journal} {Renewable Energy}\ }\textbf {\bibinfo {volume} {86}},\ \bibinfo {pages} {26--37} (\bibinfo {year} {2016})}\BibitemShut {NoStop}%
\bibitem [{\citenamefont {Kirke}(2011)}]{kirke2011}%
  \BibitemOpen
  \bibfield  {author} {\bibinfo {author} {\bibfnamefont {B.~K.}\ \bibnamefont {Kirke}},\ }\bibfield  {title} {\enquote {\bibinfo {title} {Tests on ducted and bare helical and straight blade {Darrieus} hydrokinetic turbines},}\ }\href {\doibase 10.1016/j.renene.2011.03.036} {\bibfield  {journal} {\bibinfo  {journal} {Renewable Energy}\ }\textbf {\bibinfo {volume} {36}},\ \bibinfo {pages} {3013--3022} (\bibinfo {year} {2011})}\BibitemShut {NoStop}%
\bibitem [{\citenamefont {Paraschivoiu}, \citenamefont {Trifu},\ and\ \citenamefont {Saeed}(2009)}]{paraschivoiu2009}%
  \BibitemOpen
  \bibfield  {author} {\bibinfo {author} {\bibfnamefont {I.}~\bibnamefont {Paraschivoiu}}, \bibinfo {author} {\bibfnamefont {O.}~\bibnamefont {Trifu}}, \ and\ \bibinfo {author} {\bibfnamefont {F.}~\bibnamefont {Saeed}},\ }\bibfield  {title} {\enquote {\bibinfo {title} {H-{Darrieus} {Wind} {Turbine} with {Blade} {Pitch} {Control}},}\ }\href {\doibase 10.1155/2009/505343} {\bibfield  {journal} {\bibinfo  {journal} {International Journal of Rotating Machinery}\ }\textbf {\bibinfo {volume} {2009}},\ \bibinfo {pages} {505343} (\bibinfo {year} {2009})}\BibitemShut {NoStop}%
\bibitem [{\citenamefont {Schönborn}\ and\ \citenamefont {Chantzidakis}(2007)}]{schonborn2007}%
  \BibitemOpen
  \bibfield  {author} {\bibinfo {author} {\bibfnamefont {A.}~\bibnamefont {Schönborn}}\ and\ \bibinfo {author} {\bibfnamefont {M.}~\bibnamefont {Chantzidakis}},\ }\bibfield  {title} {\enquote {\bibinfo {title} {Development of a hydraulic control mechanism for cyclic pitch marine current turbines},}\ }\href {\doibase 10.1016/j.renene.2006.02.004} {\bibfield  {journal} {\bibinfo  {journal} {Renewable Energy}\ }\textbf {\bibinfo {volume} {32}},\ \bibinfo {pages} {662--679} (\bibinfo {year} {2007})}\BibitemShut {NoStop}%
\bibitem [{\citenamefont {Gauthier}(2017)}]{gauthier2017}%
  \BibitemOpen
  \bibfield  {author} {\bibinfo {author} {\bibfnamefont {T.~A.}\ \bibnamefont {Gauthier}},\ }\emph {\bibinfo {title} {Evaluation of pitch control techniques for a cross-flow water turbine}},\ \href {https://scholarworks.montana.edu/handle/1/14070} {Ph.D. thesis},\ \bibinfo  {school} {Montana State University - Bozeman, College of Engineering} (\bibinfo {year} {2017})\BibitemShut {NoStop}%
\bibitem [{\citenamefont {Zhang}\ \emph {et~al.}(2023)\citenamefont {Zhang}, \citenamefont {Bashir}, \citenamefont {Miao}, \citenamefont {Liu}, \citenamefont {Li}, \citenamefont {Yue},\ and\ \citenamefont {Wang}}]{zhang2023}%
  \BibitemOpen
  \bibfield  {author} {\bibinfo {author} {\bibfnamefont {Q.}~\bibnamefont {Zhang}}, \bibinfo {author} {\bibfnamefont {M.}~\bibnamefont {Bashir}}, \bibinfo {author} {\bibfnamefont {W.}~\bibnamefont {Miao}}, \bibinfo {author} {\bibfnamefont {Q.}~\bibnamefont {Liu}}, \bibinfo {author} {\bibfnamefont {C.}~\bibnamefont {Li}}, \bibinfo {author} {\bibfnamefont {M.}~\bibnamefont {Yue}}, \ and\ \bibinfo {author} {\bibfnamefont {P.}~\bibnamefont {Wang}},\ }\bibfield  {title} {\enquote {\bibinfo {title} {Aerodynamic analysis of a novel pitch control strategy and parameter combination for vertical axis wind turbines},}\ }\href {\doibase 10.1016/j.renene.2023.119089} {\bibfield  {journal} {\bibinfo  {journal} {Renewable Energy}\ }\textbf {\bibinfo {volume} {216}},\ \bibinfo {pages} {119089} (\bibinfo {year} {2023})}\BibitemShut {NoStop}%
\bibitem [{\citenamefont {Le~Fouest}\ and\ \citenamefont {Mulleners}(2024)}]{lefouest2024}%
  \BibitemOpen
  \bibfield  {author} {\bibinfo {author} {\bibfnamefont {S.}~\bibnamefont {Le~Fouest}}\ and\ \bibinfo {author} {\bibfnamefont {K.}~\bibnamefont {Mulleners}},\ }\bibfield  {title} {\enquote {\bibinfo {title} {Optimal blade pitch control for enhanced vertical-axis wind turbine performance},}\ }\href {\doibase 10.1038/s41467-024-46988-0} {\bibfield  {journal} {\bibinfo  {journal} {Nature Communications}\ }\textbf {\bibinfo {volume} {15}},\ \bibinfo {pages} {2770} (\bibinfo {year} {2024})}\BibitemShut {NoStop}%
\bibitem [{\citenamefont {Strom}, \citenamefont {Brunton},\ and\ \citenamefont {Polagye}(2017)}]{strom2017}%
  \BibitemOpen
  \bibfield  {author} {\bibinfo {author} {\bibfnamefont {B.}~\bibnamefont {Strom}}, \bibinfo {author} {\bibfnamefont {S.~L.}\ \bibnamefont {Brunton}}, \ and\ \bibinfo {author} {\bibfnamefont {B.}~\bibnamefont {Polagye}},\ }\bibfield  {title} {\enquote {\bibinfo {title} {Intracycle angular velocity control of cross-flow turbines},}\ }\href {\doibase 10.1038/nenergy.2017.103} {\bibfield  {journal} {\bibinfo  {journal} {Nature Energy}\ }\textbf {\bibinfo {volume} {2}},\ \bibinfo {pages} {17103} (\bibinfo {year} {2017})}\BibitemShut {NoStop}%
\bibitem [{\citenamefont {Dave}\ \emph {et~al.}(2021)\citenamefont {Dave}, \citenamefont {Strom}, \citenamefont {Snortland}, \citenamefont {Williams}, \citenamefont {Polagye},\ and\ \citenamefont {Franck}}]{dave_aiaa_2021}%
  \BibitemOpen
  \bibfield  {author} {\bibinfo {author} {\bibfnamefont {M.}~\bibnamefont {Dave}}, \bibinfo {author} {\bibfnamefont {B.}~\bibnamefont {Strom}}, \bibinfo {author} {\bibfnamefont {A.}~\bibnamefont {Snortland}}, \bibinfo {author} {\bibfnamefont {O.}~\bibnamefont {Williams}}, \bibinfo {author} {\bibfnamefont {B.}~\bibnamefont {Polagye}}, \ and\ \bibinfo {author} {\bibfnamefont {J.~A.}\ \bibnamefont {Franck}},\ }\bibfield  {title} {\enquote {\bibinfo {title} {Simulations of {Intracycle} {Angular} {Velocity} {Control} for a {Crossflow} {Turbine}},}\ }\href {\doibase 10.2514/1.J059797} {\bibfield  {journal} {\bibinfo  {journal} {AIAA Journal}\ }\textbf {\bibinfo {volume} {59}},\ \bibinfo {pages} {812--824} (\bibinfo {year} {2021})}\BibitemShut {NoStop}%
\bibitem [{\citenamefont {Athair}\ \emph {et~al.}(2023)\citenamefont {Athair}, \citenamefont {Snortland}, \citenamefont {Scherl}, \citenamefont {Polagye},\ and\ \citenamefont {Williams}}]{athair2023}%
  \BibitemOpen
  \bibfield  {author} {\bibinfo {author} {\bibfnamefont {A.}~\bibnamefont {Athair}}, \bibinfo {author} {\bibfnamefont {A.}~\bibnamefont {Snortland}}, \bibinfo {author} {\bibfnamefont {I.}~\bibnamefont {Scherl}}, \bibinfo {author} {\bibfnamefont {B.}~\bibnamefont {Polagye}}, \ and\ \bibinfo {author} {\bibfnamefont {O.}~\bibnamefont {Williams}},\ }\bibfield  {title} {\enquote {\bibinfo {title} {Intracycle control sensitivity of cross-flow turbines},}\ }\href {https://submissions.ewtec.org/proc-ewtec/article/view/400} {\bibfield  {journal} {\bibinfo  {journal} {Proceedings of the European Wave and Tidal Energy Conference}\ }\textbf {\bibinfo {volume} {15}} (\bibinfo {year} {2023})}\BibitemShut {NoStop}%
\bibitem [{\citenamefont {Kinsey}\ and\ \citenamefont {Dumas}(2017)}]{kinsey2017}%
  \BibitemOpen
  \bibfield  {author} {\bibinfo {author} {\bibfnamefont {T.}~\bibnamefont {Kinsey}}\ and\ \bibinfo {author} {\bibfnamefont {G.}~\bibnamefont {Dumas}},\ }\bibfield  {title} {\enquote {\bibinfo {title} {Impact of channel blockage on the performance of axial and cross-flow hydrokinetic turbines},}\ }\href {\doibase 10.1016/j.renene.2016.11.021} {\bibfield  {journal} {\bibinfo  {journal} {Renewable Energy}\ }\textbf {\bibinfo {volume} {103}},\ \bibinfo {pages} {239--254} (\bibinfo {year} {2017})}\BibitemShut {NoStop}%
\bibitem [{\citenamefont {Buchner}\ \emph {et~al.}(2015)\citenamefont {Buchner}, \citenamefont {Lohry}, \citenamefont {Martinelli}, \citenamefont {Soria},\ and\ \citenamefont {Smits}}]{buchner2015}%
  \BibitemOpen
  \bibfield  {author} {\bibinfo {author} {\bibfnamefont {A.-J.}\ \bibnamefont {Buchner}}, \bibinfo {author} {\bibfnamefont {M.~W.}\ \bibnamefont {Lohry}}, \bibinfo {author} {\bibfnamefont {L.}~\bibnamefont {Martinelli}}, \bibinfo {author} {\bibfnamefont {J.}~\bibnamefont {Soria}}, \ and\ \bibinfo {author} {\bibfnamefont {A.~J.}\ \bibnamefont {Smits}},\ }\bibfield  {title} {\enquote {\bibinfo {title} {Dynamic stall in vertical axis wind turbines: {Comparing} experiments and computations},}\ }\href {\doibase 10.1016/j.jweia.2015.09.001} {\bibfield  {journal} {\bibinfo  {journal} {Journal of Wind Engineering and Industrial Aerodynamics}\ }\textbf {\bibinfo {volume} {146}},\ \bibinfo {pages} {163--171} (\bibinfo {year} {2015})}\BibitemShut {NoStop}%
\bibitem [{\citenamefont {Balduzzi}\ \emph {et~al.}(2016)\citenamefont {Balduzzi}, \citenamefont {Bianchini}, \citenamefont {Maleci}, \citenamefont {Ferrara},\ and\ \citenamefont {Ferrari}}]{balduzzi2016}%
  \BibitemOpen
  \bibfield  {author} {\bibinfo {author} {\bibfnamefont {F.}~\bibnamefont {Balduzzi}}, \bibinfo {author} {\bibfnamefont {A.}~\bibnamefont {Bianchini}}, \bibinfo {author} {\bibfnamefont {R.}~\bibnamefont {Maleci}}, \bibinfo {author} {\bibfnamefont {G.}~\bibnamefont {Ferrara}}, \ and\ \bibinfo {author} {\bibfnamefont {L.}~\bibnamefont {Ferrari}},\ }\bibfield  {title} {\enquote {\bibinfo {title} {Critical issues in the {CFD} simulation of {Darrieus} wind turbines},}\ }\href {\doibase 10.1016/j.renene.2015.06.048} {\bibfield  {journal} {\bibinfo  {journal} {Renewable Energy}\ }\textbf {\bibinfo {volume} {85}},\ \bibinfo {pages} {419--435} (\bibinfo {year} {2016})}\BibitemShut {NoStop}%
\bibitem [{\citenamefont {Dave}\ and\ \citenamefont {Franck}(2021)}]{dave_jrse_2021}%
  \BibitemOpen
  \bibfield  {author} {\bibinfo {author} {\bibfnamefont {M.}~\bibnamefont {Dave}}\ and\ \bibinfo {author} {\bibfnamefont {J.~A.}\ \bibnamefont {Franck}},\ }\bibfield  {title} {\enquote {\bibinfo {title} {Comparison of {RANS} and {LES} for a cross-flow turbine in confined and unconfined flow},}\ }\href {\doibase 10.1063/5.0066392} {\bibfield  {journal} {\bibinfo  {journal} {Journal of Renewable and Sustainable Energy}\ }\textbf {\bibinfo {volume} {13}},\ \bibinfo {pages} {064503} (\bibinfo {year} {2021})}\BibitemShut {NoStop}%
\bibitem [{\citenamefont {Strom}, \citenamefont {Johnson},\ and\ \citenamefont {Polagye}(2018)}]{strom2018SuportStruct}%
  \BibitemOpen
  \bibfield  {author} {\bibinfo {author} {\bibfnamefont {B.}~\bibnamefont {Strom}}, \bibinfo {author} {\bibfnamefont {N.}~\bibnamefont {Johnson}}, \ and\ \bibinfo {author} {\bibfnamefont {B.}~\bibnamefont {Polagye}},\ }\bibfield  {title} {\enquote {\bibinfo {title} {Impact of blade mounting structures on cross-flow turbine performance},}\ }\href {\doibase 10.1063/1.5025322} {\bibfield  {journal} {\bibinfo  {journal} {Journal of Renewable and Sustainable Energy}\ }\textbf {\bibinfo {volume} {10}},\ \bibinfo {pages} {034504} (\bibinfo {year} {2018})}\BibitemShut {NoStop}%
\bibitem [{\citenamefont {Snortland}\ \emph {et~al.}(2025)\citenamefont {Snortland}, \citenamefont {Hunt}, \citenamefont {Williams},\ and\ \citenamefont {Polagye}}]{snortland2025downstream}%
  \BibitemOpen
  \bibfield  {author} {\bibinfo {author} {\bibfnamefont {A.}~\bibnamefont {Snortland}}, \bibinfo {author} {\bibfnamefont {A.}~\bibnamefont {Hunt}}, \bibinfo {author} {\bibfnamefont {O.}~\bibnamefont {Williams}}, \ and\ \bibinfo {author} {\bibfnamefont {B.}~\bibnamefont {Polagye}},\ }\bibfield  {title} {\enquote {\bibinfo {title} {Influence of the downstream blade sweep on cross-flow turbine performance},}\ }\href {\doibase 10.1063/5.0230563} {\bibfield  {journal} {\bibinfo  {journal} {Journal of Renewable and Sustainable Energy}\ }\textbf {\bibinfo {volume} {17}},\ \bibinfo {pages} {013301} (\bibinfo {year} {2025})}\BibitemShut {NoStop}%
\end{thebibliography}%

\end{document}